\documentclass[12pt]{article}%
\usepackage[a4paper, total={7in, 8in}]{geometry}
\usepackage[utf8]{inputenc}
\usepackage[english]{babel}
\usepackage{graphicx}
\usepackage{fancyhdr}
\usepackage{amsfonts}
\usepackage{authblk}
\usepackage{mathrsfs}  
\usepackage{amsmath,amssymb,stmaryrd}
\usepackage[colorlinks=true,urlcolor=blue,linkcolor=blue]{hyperref}
\usepackage{pdfpages}
\usepackage{listings}	
\usepackage{blindtext}
\usepackage[section]{placeins}
\usepackage{nicefrac}
\usepackage[amssymb]{SIunits}
\usepackage{cite}
\usepackage{subfigure}
\usepackage{fourier}
\usepackage{caption}
\usepackage{wrapfig}
\usepackage{xcolor}
\usepackage{bbm}
\usepackage{longtable}
\usepackage{xcolor}
\usepackage{ulem}
\usepackage{url}
\usepackage{listings}
\usepackage{amsmath}
\usepackage{multirow}
\usepackage{amssymb}%
\usepackage{tikz}
\usepackage {graphicx}
\usetikzlibrary{shapes.geometric,arrows,}
\usepackage[T1]{fontenc}
\graphicspath{{figure/}}
\providecommand{\U}[1]{\protect\rule{.1in}{.1in}}

\usepackage{titlesec}
\usepackage{float}
\usepackage[textwidth=2.5cm, textsize=small]{todonotes}
\usetikzlibrary{positioning}
\usetikzlibrary{babel}
\usepackage{circuitikz}
\usetikzlibrary{babel}
\ctikzset{bipoles/resistor/height=.3}
\ctikzset{bipoles/resistor/width=0.3}

\title{Electro-optic properties of ZrO$_{2}$, HfO$_{2}$ and LiNbO$_{3}$ ferroelectric phases: A comprehensive and comparative study with density functional theory}
\makeatletter
\newcommand\email[2][]%
{\newaffiltrue\let\AB@blk@and\AB@pand
      \if\relax#1\relax\def\AB@note{\AB@thenote}\else\def\AB@note{\relax}%
        \setcounter{Maxaffil}{0}\fi
      \begingroup
        \let\protect\@unexpandable@protect
        \def\thanks{\protect\thanks}\def\footnote{\protect\footnote}%
        \@temptokena=\expandafter{\AB@authors}%
        {\def\\{\protect\\\protect\Affilfont}\xdef\AB@temp{#2}}%
         \xdef\AB@authors{\the\@temptokena\AB@las\AB@au@str
         \protect\\[\affilsep]\protect\Affilfont\AB@temp}%
         \gdef\AB@las{}\gdef\AB@au@str{}%
        {\def\\{, \ignorespaces}\xdef\AB@temp{#2}}%
        \@temptokena=\expandafter{\AB@affillist}%
        \xdef\AB@affillist{\the\@temptokena \AB@affilsep
          \AB@affilnote{}\protect\Affilfont\AB@temp}%
      \endgroup
       \let\AB@affilsep\AB@affilsepx
}
\makeatother
\author[1]{Ali El Boutaybi}
\author[2]{Panagiotis Karamanis}
\author[1]{Thomas Maroutian}
\author[1]{Sylvia Matzen}
\author[1]{Laurent Vivien}
\author[1]{Philippe Lecoeur}
\author[2]{Michel Rérat}
\affil [1]{Centre de Nanosciences et de Nanotechnologies (C2N), Universite Paris-Saclay, CNRS, 91120 Palaiseau, France}

\affil [2]{Institut des Sciences Analytiques et de Physico-Chimie pour l’Environnement et les Matériaux, CNRS, Université de Pau et des Pays de l’Adour, 64053 Pau, France}
\date{16 September 2022}
\email{\url{ali.el-boutaybi@c2n.upsaclay.fr}}

\begin{document}
\maketitle
\abstract
We report the Pockels electro-optic properties of ZrO$_2$ and HfO$_2$ orthorhombic Pbc2$_1$ and rhombohedral R3m ferroelectric phases, and we compare them to the well-known LiNbO$_3$ Pockels material from density functional theory calculations using the CRYSTAL suite of program. Specifically, three essential processes are explicitly investigated: The electronic, the ionic (or vibrational), and the piezoelectric contributions. Our calculations reveal that the ionic part coming from the low-frequency phonon modes contributes the most to the electro-optic coefficients of rhombohedral LiNbO$_3$ and of orthorhombic ZrO$_2$ and HfO$_2$. While these low-frequency modes show zero contribution to Pockels coefficients for the rhombohedral phase of the latter compounds.

\section{Introduction} \label{s:intro}
Silicon photonics platforms are becoming indispensable for developing low-cost integrated photonic circuits. Electro-optic (EO) effect is one of the main properties required in plethora applications including optical communications \cite {Mohapatra2008, Melikyan2014, Rueda2019, Abel2019}, quantum computing \cite{OBrien2009, Wang2020}, and neuromorphic applications \cite {George2019, Shen2017, Offrein2020}. The principal physical process  behind these applications is the so-called Pockels or linear EO effect, which is related to the modulation of the refractive index of a material under an applied external electric field \cite {Lines2001}. Nevertheless,  due to its centrosymmetric atomistic crystalline structure, pristine silicon features vanishing second-order optical nonlinearities; thus, it is unsuitable for EO applications. Bearing in mind that second-order nonlinear optical (NLO) responses strictly manifest in non-centrosymmetric materials, ferroelectric (FE) crystalline solids have emerged as the main materials of choice in EO applications. Such materials exhibit an intrinsic electric polarization that, in addition, can be controlled and tuned in a reversible manner by the application of an external electric field. \\

In the realm of EO applications, the most studied FE materials are currently ABO$_3$ perovskite-type  crystalline solids.  A representative member of this family is lithium niobate (LiNbO$_3$, LNO), a ferroelectric material featuring a strong Pockels crystalline bulk coefficient (r$_{33}$) of about 30 pm/V \cite{Turner1966}. LNO is widely applied in  the telecommunications industry \cite {Wooten2000-reviewonLNO} and specifically in the fabrication of optical modulators. Despite LNO's complexity of integration on silicon substrates, several studies showed its compatibility with silicon photonics \cite {Wang2018, Rabiei2013HeterogeneousLN}. An additional appealing perovskite-type crystalline solid for EO applications is tetragonal ferroelectric barium titanate (BaTiO$_3$, BTO), featuring one of the highest Pockels bulk coefficients (r$_{42}$ $\approx$ 1300 pm/V) \cite {PRB-Zgonik1994}. However, the growth of high-quality BTO thin films on silicon without affecting its strong Pockels coefficient is still challenging and of high cost, although high-quality BTO films can be fabricated via molecular beam epitaxy (MBE), \cite {Abel2019, Messner2019, Xiong2014}. In addition to tetragonal BTO,  its rhombohedral R3m phase also gained attention for cryogenic technologies as, for instance, for quantum computing \cite {Paoletta-PRB-EO, Eltes2020}.

Recently, it has been demonstrated that among HfO$_{2}$-ZrO$_{2}$-based thin films, orthorhombic Pbc2$_1$  \cite {Boscke2011} and rhombohedral R3m \cite {Wei2018} phases exhibit ferroelectric behavior. This is a significant outcome in terms of practical applications because both HfO$_{2}$ and ZrO$_{2}$ have an excellent silicon compatibility \cite {8423435-Ali,doi:10.1063/1.3636417-Muller, Park-energystorage2014, Silva-energystorage2021}. Combined with their ferroelectric nature, this might lead to significant breakthroughs in the realm of integrated photonic circuits that could significantly reduce the fabrication cost. Nevertheless, there is still much work to be done since the film quality and the ferroelectricity of ZrO$_2$ and HfO$_2$ are not well optimized at high film thicknesses. This is critical for their optimal application due to the inverse proportionality between the optical losses and the film thickness \cite {fork_armani-leplingard_kingston_1994, Epitaxialferroelectricoxide}. In the case of the orthorhombic Pbc2$_1$ (o$-$phase), which is usually reported in polycrystalline thin films, ferroelectricity has been demonstrated in a large range of thickness from 1 nm to 1$\mu$m \cite {Cheema2020, undopedHfO2, Starschich2017, Shimura-2021}. This o$-$phase has also been reported in epitaxial thin films with a thickness thinner than 20 nm \cite {Lyu2018, o-HfO2-2021, Yun2022}. Regarding the rhombohedral R3m  (r$-$phase), the reported thicknesses are ranged from 5 to 40 nm \cite {Wei2018, Silva2021, Ali2021}. Indeed, for applications, ultrathin films are often advantageous; for example, in the case of nonvolatile memories and ferroelectric field-effect transistors \cite {8423435-Ali, doi:10.1063/1.3636417-Muller}, energy storage \cite{Park-energystorage2014, Silva-energystorage2021}, negative capacitance \cite {Hoffman-negative-c, Hoffman2019}, and tunnel junctions \cite {Wei2019, Goh-2018-tj}.

In terms of NLO properties, the studies on ZrO$_2$-HfO$_2$ materials are scarce, leaving open the question of the potential of these compounds for optical applications. Indeed, to our knowledge, only two experimental studies can be found in the literature \cite {Kondo_2021_1, Kondo_2021_2}. They investigate the EO effect in Y$_2$O$_3$-doped HfO$_2$, reporting an EO coefficient of about 0.67 pm/V for (111)-oriented thin-films, a modest value compared to the ones in LNO and BTO materials. However, the exact EO coefficients of the two experimentally reported ferroelectric phases in HfO$_2$ and ZrO$_2$ are not known yet. Therefore, this study's primary objective is to report and investigate the EO coefficients of  HfO$_2$ and ZrO$_2$ materials.\\

In this work, we study the second-order nonlinear optical (NLO) properties of the orthorhombic Pbc2$_1$ and rhombohedral R3m phases of ZrO$_2$ and HfO$_2$, relying on all-electron density functional theory (DFT) computations to report reliable EO coefficients. The obtained results are compared to the ones computed  for the rhombohedral R3c LiNbO$_3$ (LNO) phase, for which experimental data are available allowing us to use LNO as a reference material. All reported EO coefficients have been determined from the microscopic second-order NLO responses of these systems that, in turn, they have been obtained via a coupled perturbed Kohn-Sham (CPKS) analytical approach developed and implemented in the CRYSTAL17 suite of quantum chemical programs by one of the authors of this study \cite{crystal14-2014, CRYSTAL17}. It has already been established that this perturbative approach delivers high accuracy molecular and bulk electric properties, free from numerical errors \cite{Maschio2012, Maschio2012-IR,  Maschio2013, Maschio2015, Rerat-2016}. To achieve the maximum possible relevance with the available or future experimental data, the reported coefficients comprise electronic, vibrational effects, and piezoelectric contributions, which have also been obtained analytically. Especially for comparison between theory and experiment, such contributions are of essential importance for a reliable prediction of the overall NLO response of a given crystalline or molecular material and the complete understanding of its origin.\\
Our article is organized as follows: First, we give the theoretical background of our study, insisting on the close relationship between the macroscopic NLO susceptibilities and the microscopic hyper-polarizabilities of the unit shells. Then, we study the EO properties of LNO using CRYSTAL code, which will serve as the material of reference. Specifically, we report reliable values of its microscopic second-order responses and analyze  the electronic, vibrational, and piezoelectric contributions. Finally, we report and discuss the Pockels EO coefficients of o$-$ and r$-$phase ZrO$_{2}$ and HfO$_{2}$, comparing them to those calculated for the R3c rhombohedral phase of LNO. 

\section{Theoretical framework and computational details} \label{s: methode}
The microscopic linear and non$-$linear optical properties of a finite zero-dimensional system  are defined in terms of a Taylor power series expansion with respect to an external applied electric field  E$_{jkl}$ as
\begin{equation}\label{eq:1}
     \mu_{i} = \mu_{0} + \alpha^{\textit{tot}}_{ij} E_{j}+ \frac{1}{2} \beta^{\textit{tot}}_{ijk} E_{j} E_{k} +  \frac{1}{6} \gamma^{\textit{tot}}_{ijkl} E_{j} E_{k}E_{l}  +...
\end{equation}
Where, \textit{i}, \textit{j}, \textit{k} and l stand for Cartesian coordinates (\textit{x}, \textit{y}, \textit{z}) ,  $\mu_{i}$ and $\mu_{0}$ are the induced and the permanent dipole moment,  $\alpha^{\textit{tot}}_{ij}$ is the total dipole electric polarizability, while $ \beta^{\textit{tot}}_{ijk}$ and  $\gamma^{\textit{tot}}_{ijkl}$ are the total first and second dipole hyperpolarizabilities, respectively. By the term total (\textit{tot}) it is implied that both electronic (\textit{ele}) and vibrational (\textit{vib}) contributions\cite{Bishop1990, Maschio2015, Rerat-2016}  are taken into account, Hence:
\begin{equation}\label{eq:2}
 \alpha^{\textit{tot}}_{ij} =  \alpha^{ele}_{ij} +\alpha^{vib}_{ij}
\end{equation}
\begin{equation}\label{eq:3}
  \beta^{\textit{tot}}_{ijk} = \beta^{ele}_{ijk} + \beta^{vib}_{ijk}
\end{equation}
For three-dimensional periodic crystalline lattices,  equation (\ref{eq:1}) can be written as
\begin{equation}\label{eq:4}
    P_{i} = P_{0} + \chi^{(1)}_{ij} E_{j} +  \chi^{(2)}_{ijk} E_{j}  E_{k} + \chi^{(3)}_{ijkl} E_{j}  E_{k} E_{l}+...
\end{equation}\\
Where $P_{i}$ is the unit-cell induced polarization while   $\chi^{(1)}_{ij}$, $\chi^{(2)}_{ijk}$,  and $\chi^{(3)}_{ijkl}$ stand for the total first order (linear), and the second-and the third-order (nonlinear) optical macroscopic bulk susceptibilities, respectively. The first and second coefficients of equation (\ref{eq:4}) are deduced from the total microscopic (hyper) polarizabilities per unit-cell volume (\textit{V}) as follows (in atomic units):

\begin{equation}\label{eq:5}
 \chi^{(1)}_{ij} =  \frac {4 \pi}{V}\alpha^{\textit{tot}}_{ij}
\end{equation}

\begin{equation}\label{eq:6}
  \chi^{(2)}_{ijk} =   \frac {2 \pi}{V} \beta^{tot}_{ijk} 
\end{equation}
The total first-order susceptibility $\chi^{(1)}_{ij}$ can be used to deduce the relative dielectric tensor $\epsilon_{ij}$, which in turn gives access to bulk refractive indices $n$ via the following expressions: 
 
 \begin{equation}\label{eq:7}
 \epsilon_{ij} = \delta_{ij} + \chi^{(1)}_{ij} 
 \end{equation}

\begin{equation}\label{eq:8}
 n_{ii} = \epsilon_{ii}^{1/2}   
 \end{equation}
In the above equations, $\delta_{ij}$ represents the elements of the identity matrix ($\delta_{ij}=1$ for $i=j$, and $0$ for $i\ne j$). Note that we consider in this work a diagonal $\epsilon_{ij}$ tensor so that the subscript $i$ of $\epsilon_{ii}$ corresponds to one of the three principal axes $n_{ii}$ of the ellipsoidal indicatrix. 

Now, to obtain the EO Pockels coefficients ($r_{ijk}$), as a first step, we consider the constant stress (or "clamped") conditions, wherein $\chi^{(2)}_{ijk}$ and the bulk refractive indices $n_{ii}$ given in equation (\ref{eq:8}) can be used to determine one out of the two terms contributing to the total $r_{ijk}$ coefficients. The clamped EO coefficient is denoted here as $r^{S}_{ijk}$, comprises both vibrational and electronic contributions, and is defined  as:
\begin{equation}\label{eq:9}
   r^{S}_{ijk} =  - 2 \,\,\,   \frac {\chi^{(2)}_{ijk}}{n^{2}_{ii} n^{2}_{jj}}
\end{equation}

In the second step, we should model and consider the relevant acoustic contribution stemming from the piezoelectric effect. This effect depends on the frequency of the applied ac electric field, as it originates from the inverse piezoelectric effect, according to which the application of an ac oscillating field induces mechanical deformations on the crystalline lattice. The resulting contribution to the $r_{ijk}$ coefficients, denoted here as $r^{p}_{ijk}$, is defined from  the photoelastic tensor  $P_{ij\mu \nu}$ and converse piezoelectric tensor $d_{k \mu \nu}$ \cite {Ebra2015, Zgonik2002} as:
\begin{equation}\label{eq:10}
  r^{p}_{ijk} =   \sum_{\mu , \nu =1}^{3} P_{ij\mu \nu} d_{k\mu \nu}    
\end{equation}
The calculation of the tensorial components of $r^{p}_{ijk}$ is achieved via the computation of the photoelastic constants $\frac {\partial \eta_{ij}}{\partial S_{\mu \nu}}$ ($P_{ij\mu \nu}$) defined as the derivative of
the impermeability tensor $\eta_{ij}$ with respect to strain $S_{\mu \nu}$, which are the elements of the fourth rank photoelastic tensor, and of the converse piezoelectric tensor $\frac {\partial S_{\mu \nu}}{\partial E_{k}}$ ($d_{k\mu \nu}$)  \cite {PRB-Ebra2013}. Finally, the total EO coefficients that we report in this work are calculated as $r_{ijk} =r^{S}_{ijk} + r^{p}_{ijk}$. More detailed information about photoelastic calculations can be found elsewhere\cite {PRB-Ebra2013, Ebra2015, Zgonik2002}. \\

All computations of frequency-dependent properties have been performed with the CRYSTAL suite of programs \cite {CRYSTAL17}. In order to determine the dynamic vibrational  and electronic (clamped nuclei) hyper-polarizabilities, a fully analytical approach relying on the coupled perturbed Hartree-Fock and Kohn-Sham method (CPHF / CPKS) has been applied \cite{Maschio2015, Rerat-2016, Pascale2004, Zicovich2004, Maschio2013}. 
In the case of  LNO, we considered three different DFT functionals to investigate the dependence on the DFT method of the properties of interest and to compare the results with available experimental data. That includes the Becke, three-parameter, Lee-Yang–Parr exchange-correlation functional (B3LYP) \cite {Becke1993, Lee1988}, the pure Perdew-Burke-Ernzerhof one (PBE) \cite {PBE1996}, and with 25$\%$ Hartree-Fock (HF) exchange (PBE0) \cite {Adamo1999}. In the case of   ZrO$_{2}$ and HfO$_{2}$ (orthorhombic Pbc2$_1$ and rhombohedral R3m phases), only the B3LYP functional has been applied as it gives good results for LNO. The basis sets of LNO are taken from Refs. \cite {PRB-basisset-Nb}, \cite {Oliveira-basisset-Li}, and \cite {PRB-basisset-O} for Nb, Li, and O, respectively. The basis sets for ZrO$_{2}$ are from Refs. \cite {Oliveira-basisset-Li} and \cite {Valenzano2011} for O and Zr, respectively, while those for HfO$_2$ are taken from Refs. \cite{Basisset-O-HfO2} and \cite {PRB-basisset-Hf} for O and Hf, respectively.
\section{Results and discussion} \label{s: discussion}

\begin{figure}[H]
    \centering
     \captionsetup{width=.8\linewidth}
     \hspace*{-0.5in}
    \includegraphics [scale=0.6] {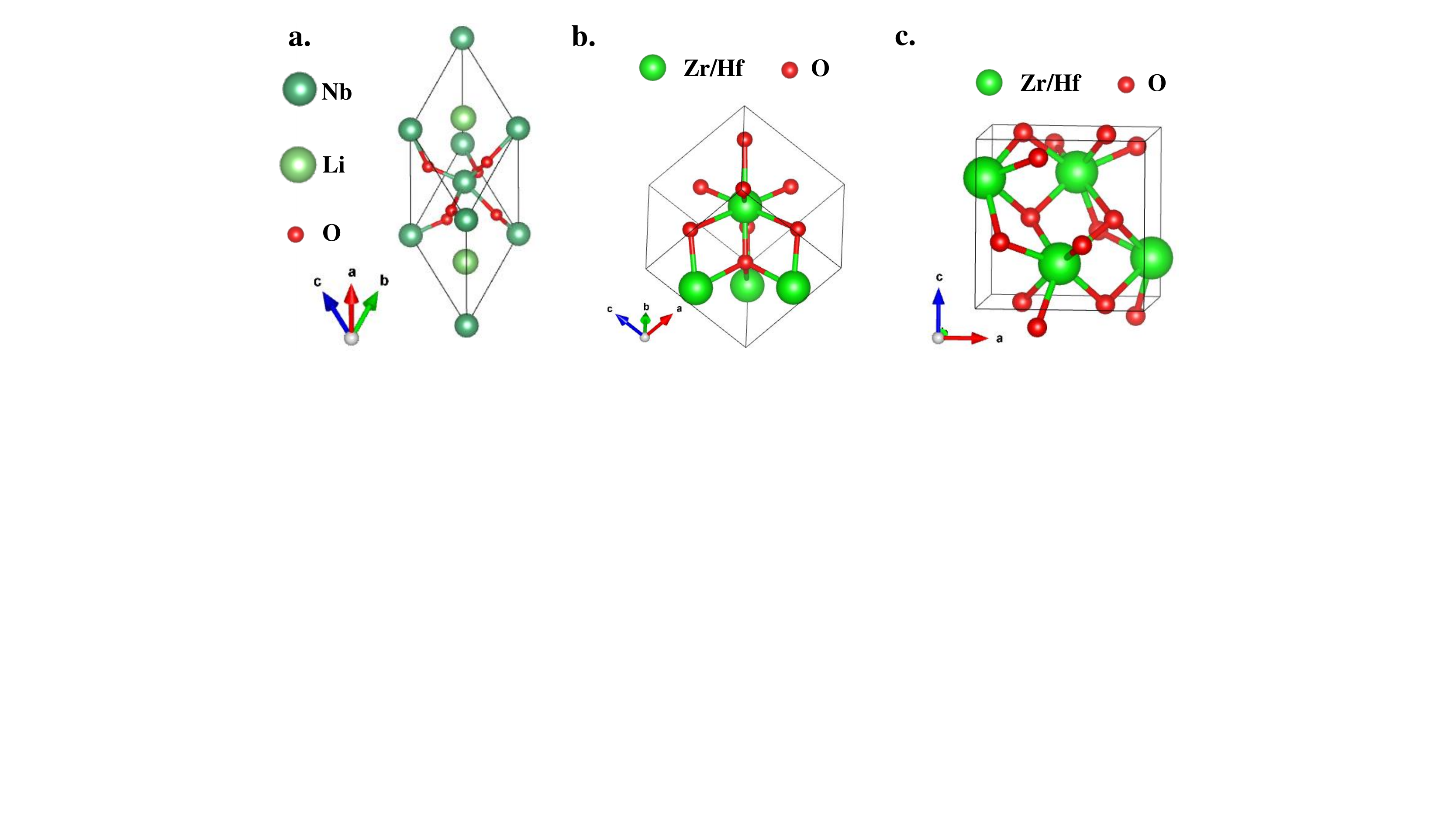}
    \vspace*{-70mm}
    \caption{The crystallographic structures of a) LiNbO$_3$ (space group R3c), b) rhombohedral ZrO$_2$-HfO$_2$ (space group R3m), and c) orthorhombic ZrO$_2$-HfO$_2$ (space group Pbc2$_1$).}
    \label{fig:1} 
\end{figure}

\subsection{LNO rhombohedral R3c} \label{ss:LiNbO3}
We started our calculations with LNO, which features a ferroelectric R3c ground state at low temperatures. The crystal structure of the bulk system, built from ten atoms per unit cell, is shown in Figure \ref{fig:1}a (lattice constants and atomic  coordinates are given in Table S.1 in Supplemental Material (SM) \cite{Ali2022}). Computed  band gaps,  dynamic refractive indices n$_{ii}$ and $\epsilon_{ii}$  values of LNO determined  with three functionals of different types, namely B3LYP, PBE0, and PBE at an optical frequency corresponding to 633 nm wavelength, are listed in  Table \ref{table:1}. Apart from  electronic contributions, which can be trivially computed with the CPKS approach, the reported values comprise as well the respective vibrational contributions computed at the same level of theory. The wavelength (633 nm) was chosen to compare our results to available data reported in the literature; both experimental and theoretical data are reported in Table \ref{table:1}. Note that the EO properties are also investigated up to the near-infrared wavelength range in section \ref{ss:o-ZHO}.

Starting from the computed bandgap of LNO, it is seen that out of the three methods used, B3LYP and PBE0 yield notably larger values as compared to the reported experimental values. On the other hand, the pure PBE functional follows the opposite trend. The frequently cited direct band gap value of LNO, mostly concluded from optical experiments \cite {GapLNO-1990}, is about 3.78 eV. However, this value should be considered with care due to electron-hole attraction effects, which might induce important underestimations in the final outcome of optical measurement, as already discussed in Refs.  \cite {Thierfelde-gap2009, PRB-Nahm2008}. Therefore, for LNO crystal, a band gap of 4.7 eV \cite {Thierfelde-gap2009} should be considered more suitable for comparisons between theory and experiment. Bearing in mind the latter correction, the B3LYP functional and the associated basis set reproduce the band gap of LNO better than PBE.

Turning our attention to the refractive indices computed in this study, we also notice that both B3LYP and PBE0 functionals yield values that fall very close to the available experimental measurements. On the other hand,  pure PBE overshoots the experimental reference values, yielding values close to the earlier Local-Density approximations (LDA) reported results. This behavior is not surprising since pure DFT functionals generally underestimate the gap returning overestimated cell polarizabilities $\alpha_{ij}$ \cite {Bishop1990} and  dielectric constants. Hence, it is  expected to affect the refractive indices (equation \ref {eq:8}). Nonetheless, despite the apparent deviations in functional performance, all methods considered predict relatively stable birefringence values (denoted $ \delta n$ in Table \ref{table:1}), in good agreement with the available experimental measurements conducted on stoichiometric LNO crystals (cationic ratio Li/Nb $\sim$ 1) \cite{Schlarb1993, Bergman1968}. 

\begin{table} [H]
    \centering
    \captionsetup{width=.8\linewidth}
     \caption{Band gap, refractive indices $n_{ii}$, the birefringence $ \delta n$, and dielectric constants $\epsilon_{ii}$ of LNO  calculated at 633 nm wavelength using different functionals and compared to experimental results. Values in parentheses correspond to the static values (at "infinite" wavelength).}
\begin{tabular}{  c c c c c c c c} 
\hline
&B3LYP & PBE0&  PBE&LDA \cite {PRBVeithen2004}& exp \cite{Schlarb1993} &   exp \cite{ Nelson1974}\\ [1ex] 
\hline\hline
Gap (eV) &4.75& 5.29& 3.15& -&   3.78  &-\\[0.5ex] 
n$_{xx}$ (n$_{o}$) &2.257&2.227 &2.497& 2.37 & 2.28 & 2.286\\[0.5ex] 
n$_{zz}$ (n$_{e}$)& 2.149&2.120& 2.402& 2.35 &  2.18& 2.202\\[0.5ex] 
$ \delta n$& 0.107& 0.107&  0.095&/ &  0.098& 0.09\\[0.5ex] 
$\epsilon_{xx}$ &5.10 (4.71) & 4.96& 6.23 &/ &   / &/\\[0.5ex] 
$\epsilon_{zz}$ &4.63 (4.31)& 4.50& 5.77& / &   /  &/\\[0.5ex] 
  \hline 
\end{tabular}
 \label{table:1}
\end{table}

The clamped EO coefficients  r$^{S}_{ijk}$ of LNO, together with the corresponding electronic and vibrational contributions computed with the B3LYP functional, are listed in Table \ref{table:2}. We also list previous  published experimental measurements and theoretical data in the same table. All properties have been deducted from computed $\chi^{(2)}_{ijk}(-\omega_{\sigma}; \omega, 0)$ (for Pockels coefficient calculation, $\omega$ stands for a laser wavelength, $\omega = 0$ for a static electric field, and  $\omega_{\sigma}$ is the sum of two wavelengths, see  SM for more details \cite{Ali2022}) obtained at the same level of theory, considering only the four EO independent tensorial components of LNO  r$_{33}$, r$_{13}$, r$_{11}$, and r$_{51}$ \cite {PRBVeithen2004}. The vibrational contribution is a result of the transverse optical phonon modes; they can be represented as $4A_1+ 5A_2 + 9 E$, where the $A_1$ and $E$ modes are Raman and infrared (IR) active; in contrast, $A_2$ modes are silent. The $A_1$ modes are coupled to r$_{33}$ and r$_{13}$, while the $E$ modes are linked to r$_{11}$ and r$_{51}$. The corresponding computed phonon frequencies, showing a good agreement with available experimental data, together with the second-order nonlinear susceptibilities $\chi^{(2)}_{ijk}(-\omega_{\sigma}; \omega,0)$ at a wavelength of 633 nm, are given in SM \cite{Ali2022}. Looking at the computed clamped EO coefficients presented in Table \ref{table:2}, the results obtained with B3LYP functional are in very good agreement with experimental results, notably for r$_{33}$ and r$_{13}$ coefficients. On the other hand, an overestimation of r$_{11}$ and an underestimation of r$_{51}$, respectively, are revealed. Nevertheless, our calculations expose that the  most dominant contribution out of the two considered for r$^{S}_{ijk}$ is the vibrational contribution to the microscopic second hyperpolarizability of the unit cell. This result, relying on a hybrid DFT treatment of the wavefunction, confirms previous computations within the LDA approximation \cite {PRBVeithen2004}, producing at the same time a very good estimation of the EO tensorial components with respect to the experimental data.

\begin{table} [H]
    \centering
    \captionsetup{width=.8\linewidth}
     \caption{ $r_{ijk}(-\omega_{\sigma}; \omega, 0)$ EO tensor (pm/V) of LNO at 633 nm wavelength: electronic and ionic (clamped), and piezoelectric (unclamped) contributions, together with results from previous calculations and experiments.}
\begin{tabular}{c c c c c c c c c c} 
\hline
\multirow{2}{1.5cm} &  &\multicolumn{3}{p{3cm}}{\centering B3LYP}  &  \multicolumn{1}{p{1.6cm}}{\centering LDA \cite {PRBVeithen2004}} &   \multicolumn{1}{p{1.5cm}}{\centering exp \cite{Turner1966}} &  \multicolumn{1}{p{1.5cm}}{\centering exp \cite {K.K.WONG2002}}\\ [1ex]
\hline
Clamped & & $r^{ele}_{ijk}$& $r^{vib}_{ijk}$& $r^{S}_{ijk}$&\\ [1ex]
 &r$_{33}$ &    7.19& 23.45 &30.64& 26.9   & 30.8  &34\\  
&r$_{13}$ &   2.74 & 7.94& 10.68  &9.7& 8.6 &10.9\\  
&r$_{11}$ & -1.02 &    -6.84 & -7.86& 4.6  & 3.4 &/\\  
&r$_{51}$ & 3.08&    19.43 & 22.51& 14.9 &  28&/\\  
\\
Unclamped & &$r^{p}_{ijk}$ &  &  $r_{ijk}$ &  & &exp \cite{Abdi1998}\\[1ex]
& r$_{33}$ & 0.95 &     &31.59&  27   &32.6 & /\\  
&r$_{13}$ & 1.42 &   &12.1&  10.5  &10& /\\  
&r$_{11}$ & 0.86 &   & -7.00&   7.5  &6& 9.9\\  
&r$_{51}$ & 10.77 &   & 33.28&  28.6& 32.2& /\\  
\hline  \hline
\end{tabular}
\label{table:2}
\end{table}

We now consider the total unclamped EO coefficients $r_{ijk}$ of LNO, including the piezoelectric contribution, which is computed at the same level of theory as discussed above. We note that the photoelastic constants $P_{ij\mu \nu}$, here computed at 633 nm, are found relatively independent on wavelength above 633 nm (see Figure S.2 in SM \cite{Ali2022}). The same trend has been observed for MgO and CaWO$_4$ crystals from calculations at the PBE level \cite{PRB-Ebra2013, Ebra2015}. The piezoelectric contribution to the EO coefficients is listed in Table \ref{table:2} as $r^{P}_{ijk}$. Looking at the total $r_{ijk} =r^{S}_{ijk} + r^{p}_{ijk}$, an excellent agreement with experimental results is obtained. For the three coefficients r$_{33}$, r$_{13}$, and r$_{11}$, the vibrational contribution is still dominant, while r$_{51}$ coefficient has the highest piezoelectric contribution (Table \ref{table:2}). Three functionals, B3LYP (Table \ref{table:2}), PBE0, and PBE (Table S.5), give a piezoelectric contribution to r$_{51}$ of around 11 pm/V, that represents around 33 $\%$ of  r$_{51}$ unclamped value, lower than the 48 $\%$ found in Ref.\cite {PRBVeithen2004}. Nevertheless, the r$_{51}$ EO coefficient is found with the highest piezoelectric contribution from the four independent EO coefficients of LNO material, in line with the previous theoretical results \cite {PRBVeithen2004}. Experimentally, only a piezoelectric contribution of about 4.2 pm/V has been reported \cite{Turner1966}. This disagreement might be due to the experimental method, which can not suppress the total piezoelectric contribution during the measurement of the clamped coefficient, as previously suggested in Ref.\cite {PRBVeithen2004}.
 
It is worth noting that the PBE0 functional showed an overestimation of the piezoelectric contribution to r$_{33}$ ($\sim$6 pm/V see Table S.5), which is higher than the value obtained with B3LYP and PBE functionals ($\sim$1 pm/V close to the 1.8 pm/V found experimentally  \cite{Turner1966}). Moreover, pure PBE functional delivers a reasonable estimation of the EO coefficients (Table S.5), even though it underestimates the gap, thus, giving a high refractive index (Table \ref{table:1}) as discussed above. That is because underestimated gaps lead to overestimated susceptibility $\chi^{(2)}_{ijk}$; in return, this will compensate for the high values of the refractive indices when computing the EO coefficients r$^{S}_{ijk}$ (see equation \ref {eq:5}).

\subsection{ZrO$_{2}$-HfO$_{2}$ orthorhombic Pbc2$_{1}$} \label{ss:o-ZHO}
In this section, we present electro-optics properties of the o$-$phase of ZrO$_{2}$ and HfO$_{2}$ compounds, with space group Pbc2$_{1}$ (No. 29). The optimization of Pbc2$_{1}$ ZrO$_{2}$ and HfO$_{2}$ is performed using 12 atoms in a unit cell. The optimized crystalline structure and atomic positions are given in SM \cite {Ali2022}. The polarization was also computed for o$-$phase ZrO$_{2}$ and HfO$_{2}$ via Berry phase calculation using Quantum Espresso (QE) \cite {Giannozzi2009}, in order to verify that the optimized polar phases are consistent with the ones reported in literature \cite {Boscke2011, Materlik2015}. Table \ref {table:3} displays the band gap, refractive indices, and remanent polarization of o$-$phase ZrO$_{2}$ and HfO$_{2}$; Note that regarding optical properties, no experimental data are available for comparison. However, we did obtain polarization values similar to previous calculations \cite {Materlik2015}.\\
It is well known that ZrO$_{2}$ and HfO$_{2}$ compounds have a wide band gap, as they are used for high-k applications. From Table \ref {table:3} it can be seen that the band gap of o-HfO$_2$ is higher by about 1 eV compared to o-ZrO$_2$. Another feature is that the refractive indices at 633 nm of o-ZrO$_{2}$ are higher than the ones of o-HfO$_{2}$. This is interesting because a high refractive index is an essential element for silicon photonics applications in order to ensure a sufficient contrast with the refractive index of SiO$_2$ \cite {Messner2019}. 
\begin{table} [H]
    \centering
    \captionsetup{width=.6\linewidth}
     \caption{ Band gap and refractive indices of  orthorhombic Pbc2$_1$  ZrO$_{2}$ and HfO$_{2}$  at 633 nm. The polarization P$_b$ (along the b axis) from Berry phase calculation is also shown.}
\begin{tabular}{c c c} 
 \hline
   & o-ZrO$_{2}$ & o-HfO$_{2}$\\ [1ex] 
 \hline\hline
 gap (eV)&5.40& 6.45\\  [1ex]
 n$_{xx}$ & 2.204& 2.059 \\
 n$_{yy}$ & 2.301  & 2.122 \\  
 n$_{zz}$ & 2.236 & 2.081  \\  [1ex] 
 P$_b$ ($\mu$C/cm$^{2})$ & 54 (58, \cite {Materlik2015} ) &49.5 (50, \cite {Materlik2015} ) \\ 
 \hline  \hline 
\end{tabular}
 \label{table:3} 
\end{table} 
We also have to consider the vibrational modes of the Pbc2$_{1}$ phase, as these modes are directly linked and contribute to the EO coefficients. The primitive unit cell of Pbc2$_{1}$ phase with 12 atoms (4 formula units) results in 36 vibrational modes, 3 for translation modes, and 33 optical modes, having the following irreducible representation at the zone center $ \Gamma$: \\
\begin{equation}\label{eq:11}
 \Gamma = 8 A_{1} + 9 A_{2} + 8 B_{1} + 8 B_{2}
\end{equation}
The  $A_{1}$, $B_{1}$, and $B_{2}$ modes are Raman and IR active. Hence, they will contribute to the electro-optic coefficients, while $A_2$ modes are only Raman active. The calculated $A_1$ phonon frequencies are given in Table \ref{table:4}, and the $A_{2}$, $B_{1}$, and $B_{2}$ phonon frequencies are given in SM \cite {Ali2022}. Only a few experimental results concerning the phonon frequencies of Pbc2$_{1}$ o$-$phase are available for pure ZrO$_2$ and HfO$_2$. They are displayed for o-ZrO$_2$ in Table \ref{table:4}, showing a good agreement with the calculated phonon modes.\\ 
\begin{table}[H]
    \centering
    \captionsetup{width=.7\linewidth}
     \caption{Calculated frequencies ($cm^{-1}$) of Raman and IR active $A_1$ phonon modes of ferroelectric ZrO$_{2}$ and HfO$_{2}$ using B3LYP functional.}
\begin{tabular}{c c c c c c c c c c}
\hline
\multirow{2}{1.5cm} &  \multicolumn{2}{p{2.5cm}}{\centering o-ZrO$_2$}  &  \multicolumn{1}{p{1.5cm}}{\centering o-HfO$_2$} &   \multicolumn{1}{p{1.5cm}}{\centering r-ZrO$_2$} &  \multicolumn{1}{p{1.5cm}}{\centering r-HfO$_2$}\\   [1ex] 
  & \multicolumn{1}{c}{} & \multicolumn{1}{c}{exp \cite {Uwe-2022-Raman}}  & \multicolumn{1}{c}{} &   \multicolumn{1}{c}{}  & \multicolumn{1}{c}{}  \\   [1ex] 
A$_1$ & 102 &   /& 113 &171 &118\\  
 & 186 &200& 153&  220&171\\  
 & 291 &  /& 253 &267&267\\  
 & 315& 320& 295&352&284\\  
 & 351& 340& 329& 449&423\\  
 &387&/ &369&534&537\\
 &445&/&451&674&692\\
 &566& 580&589&755&770\\
\hline  \hline 
\end{tabular}
\label{table:4}
\end{table}
The orthorhombic Pbc2$_{1}$ with a 2mm point group has five independent EO coefficients \cite {modernphotonics}: r$_{13}$, r$_{23}$, r$_{33}$, r$_{42}$, and r$_{51}$.  Moreover, as given above, only  $A_{1}$, $B_{1}$, and $B_{2}$ contribute to the EO coefficients in the Pbc2$_1$ phase. The $A_{1}$ modes are linked to r$_{13}$, r$_{23}$, and r$_{33}$, the $B_{1}$ and $B_{2}$ modes are coupled to r$_{42}$ and r$_{51}$, respectively. These five EO coefficients are computed at the same level of theory as discussed for LNO, and they are listed in Table \ref {table:5} showing the three different contributions, electronic r$^{ele}_{ijk}$, vibrational r$^{vib}_{ijk}$, and piezoelectric r$^{p}_{ijk}$. The first significant result is that the  EO coefficients in o-ZrO$_{2}$ are higher and almost twice the ones in o-HfO$_{2}$. The r$_{13}$  and r$_{33}$ coefficients have the highest values of the five independent EO elements. For both ZrO$_{2}$ and HfO$_{2}$, these two coefficients have an opposite sign and are quite close in absolute values. Comparing the calculated EO results of o-ZrO$_2$ to those of the reference LNO material (Table \ref {table:2}), the r$_{13}$ of o-ZrO$_{2}$ is close to the r$_{13}$  of LNO, while the r$_{33}$ of o-ZrO$_{2}$ is around one-third  of LNO. Unfortunately, no experiment results are available for pure o-ZrO$_{2}$ and o-HfO$_{2}$. However, for Y$_{2}$O$_{3}$-doped HfO$_{2}$ o$-$phase, EO coefficients have been reported with effective values of about 0.46 pm/V and 0.67 pm/V for (100) and (111) oriented thin films, respectively \cite {Kondo_2021_1, Kondo_2021_2}. 

\begin{figure} [H]
    \centering
    \captionsetup{width=.6\linewidth}
     \hspace*{-0.5in}
    \includegraphics [scale=0.6]{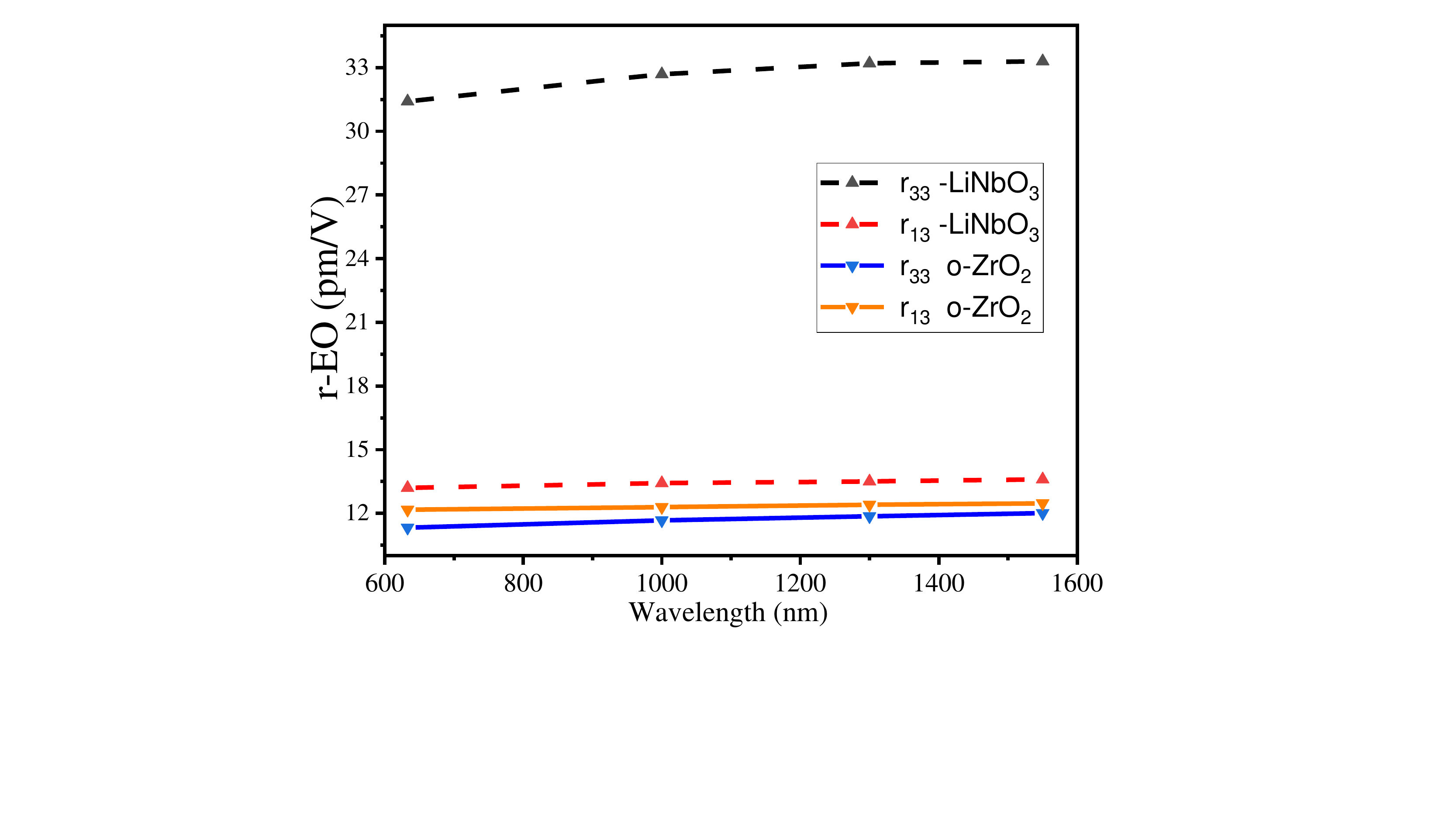}
    \vspace*{-30mm}
    \caption{EO coefficients (unclamped) r$_{33}$ and r$_{13}$ of LNO and o-ZrO$_2$ at different wavelengths using B3LYP functional.}
    \label{fig:2}
\end{figure}

\begin{figure} [H]
    \centering
     \captionsetup{width=.5\linewidth}
     \hspace*{-0.2in}
    \includegraphics [scale=0.5]{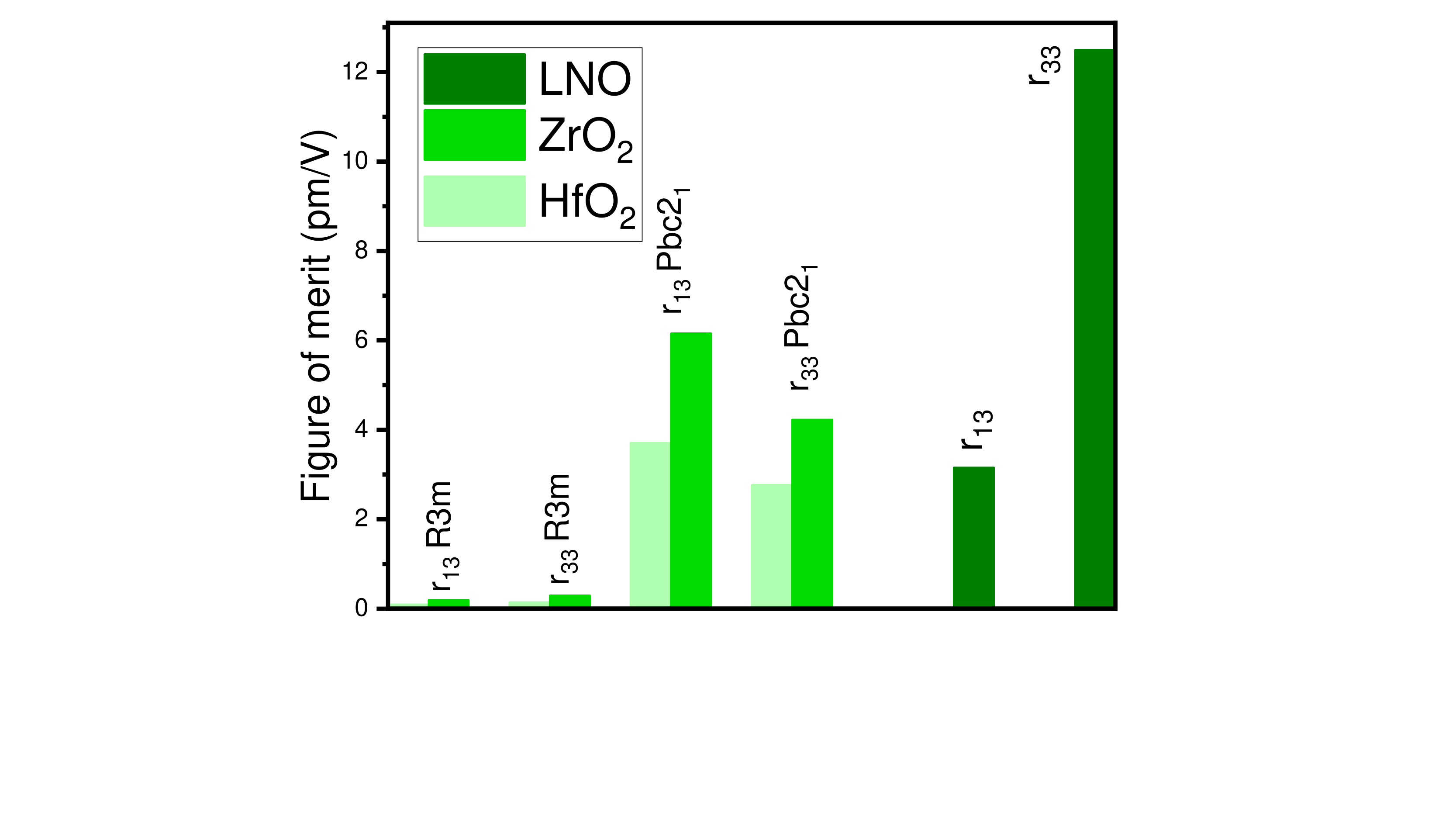}
     \vspace*{-22mm}
    \caption{Figure of merit ($n^{3}$  r$_{ijk}$/ $\epsilon$) of HfO$_2$ and ZrO$_2$ compared to LNO using  r$_{33}$ and r$_{13}$ coefficients.}
    \label{fig:3}
\end{figure}

From Table \ref{table:5}, we also deduce that ionic is the dominant contribution to the EO coefficients in o$-$phase ZrO$_2$ and HfO$_2$, as in LNO. The approach used here for EO calculations also provides further insights into this ionic contribution computed by evaluating the Raman and IR intensities \cite{Rerat-2016}, as described in refs.  \cite{Maschio2013} and \cite{Maschio2012}, respectively. In the case of LNO crystal, we found that the two $A_1$ and $E$ polar modes of lowest frequency contribute the most to r$_{33}$ and r$_{51}$, respectively, amounting to 43$\%$ and 36$\%$ of the value of these coefficients. This observation is in good agreement with previous calculations \cite {PRBVeithen2004}, and with the fact that for LNO, unstable phonon modes in the paraelectric phase transform into lower frequency and highly polar modes in the low-temperature ferroelectric phase \cite {PRBVeithen2004}. Nevertheless, ZrO$_2$ and HfO$_2$ materials have numerous polymorphs \cite{Kisi1998} and other mechanisms that can stabilize the ferroelectric phase \cite{Materlik2015, Kisi}, away from the classical paraelectric-ferroelectric phase transition picture. However, looking at the lowest phonon modes in o-ZrO$_2$, the same trend is observed as in LNO material. For example, the lowest frequency $A_1$ TO (transverse optical) mode at 102 $cm^{-1}$ (Table \ref{table:4}) is responsible for about 40$\%$ and more than 60$\%$ of the values of r$_{33}$ and r$_{13}$ in o-ZrO$_2$, respectively. 
\begin{table}[H]
    \centering
    \captionsetup{width=.8\linewidth}
     \caption{The five independent elements of $r_{ijk}(-\omega_{\sigma}; \omega, 0)$ tensor for Pbc2$_{1}$ space group, together with their respective electronic, vibrational and piezoelectric contributions, computed at a wavelength of 633 nm. All values are given in pm/V.}
\begin{tabular}{c c c c c c c c c c}
\hline
\multirow{5}{1.5cm} &  & \multicolumn{5}{p{2.5cm}}{\centering o-ZrO$_2$}  &  \multicolumn{2}{p{2.5cm}}{\centering o-HfO$_2$} \\   [1ex] 
  & \multicolumn{1}{c}{r$^{ele}_{ijk}$ } & \multicolumn{1}{c}{r$^{vib}_{ijk}$ } & \multicolumn{1}{c}{r$^{p}_{ijk}$ } & \multicolumn{1}{c}{r$_{ijk}$} & & \multicolumn{1}{c}{r$^{ele}_{ijk}$ } & \multicolumn{1}{c}{r$^{vib}_{ijk}$ } & \multicolumn{1}{c}{r$^{p}_{ijk}$ } & \multicolumn{1}{c}{r$_{ijk}$}\\   [1ex] 
 r$_{33}$ & 2.10 &   8.75& 0.5 & 11.35& & 1.52 & 5.56  &0.14&7.22 \\  
r$_{13}$ & -1.49 &- 9.29&   -1.38 &-12.17 & & -0.81 &  -6.32 & -0.83&-7.96\\  
r$_{23}$ & -0.51 &  -2.19&0 & -2.70& & -0.51 &  -1.94 & 0&-2.58 \\  
r$_{42}$ &- 0.53 &  -0.95& 0& -1.48& & -0.54 & -0.83 &   0&  -1.37\\  
r$_{51}$ & -1.00 &  -1.16& 0& -2.16& & -0.83 &-0.94&  0&  -1.77\\  
\hline  \hline 
\end{tabular}
\label{table:5}
\end{table}
\textit{Wavelength dependence of EO coefficients}: The EO values reported above are computed at 633 nm, while, for example, wavelengths for telecom applications are in the near-infrared region, typically around 1500 nm. Therefore, with a focus on the influence of laser wavelength on Pockels coefficients, the same calculations as performed at 633 nm were repeated at different wavelengths up to near-infrared. The results are displayed in Figure \ref {fig:2} for the unclamped r$_{33}$ and r$_{13}$ coefficients of LNO and o-ZrO$_2$. ZrO$_2$ is here chosen instead of HfO$_2$ because of its higher EO coefficients compared to HfO$_2$ (Table \ref{table:5}).
First, bearing in mind that the three contributions depend on wavelength, the electronic part is more sensitive to the electric field than the vibrational and piezoelectric contributions. Also, the electronic part depends strongly on the gap of the materials. Therefore, approaching the IR wavelengths, the electronic contribution decreases and converges to a static value \cite {Maschio2015, Rerat-2016}. This decrease is associated with a decrease in the refractive indices: At a wavelength of 1550 nm, the refractive index n$_{zz}$ is found to be 2.08 (2.19) for LNO (for ZrO$_{2}$), which is lower than the value obtained at 633 nm (Tables \ref{table:1} and \ref{table:3}). We found that the vibrational contribution remains almost constant from 633 nm to 1550 nm. The last contribution is piezoelectric, where the dependency on wavelength comes from the photoelastic coefficients, which remain relatively constant at high wavelengths (see Figure S1 in SM \cite{Ali2022}). The general relative increase of the total EO coefficients observed in Figure \ref {fig:2} when increasing the wavelength is thus due to the decrease of the refractive indices (see equation \ref{eq:9}). The EO coefficients of o-ZrO$_2$ are found lower than the ones of LNO at all investigated wavelengths. Nevertheless, in terms of applications, other parameters should be taken into consideration, such as the refractive index and the total dielectric constant, to define a figure of merit ($n^{3}$  r$_{ijk}$/ $\epsilon$ \cite{Epitaxialferroelectricoxide}, with $\epsilon$ a total dielectric constant comprising both the electronic and the ionic contributions) that could bring o$-$phase ZrO$_2$/HfO$_2$ as a competitive material for EO applications. Figure \ref{fig:3} displays the comparison of this figure of merit between LNO, HfO$_2$ and ZrO$_2$ materials. For r$_{33}$ coefficient, LNO has the highest figure of merit; however, the ones for ZrO$_2$ and HfO$_2$ are not negligible. Considering the compatibility of ZrO$_2$ and HfO$_2$  with silicon, these results are significant enough for further investigating different paths to increase the EO coefficients in these materials. 
\subsection{ZrO$_{2}$-HfO$_{2}$ rhombohedral R3m} \label{ss:r-ZHO}
In this section, we focus on the EO properties of the rhombohedral R3m (space group No. 160) phase of ZrO$_{2}$-HfO$_{2}$. This phase is less studied than the orthorhombic phase and has only been reported in compressively strained epitaxial thin films \cite {Wei2018, Ali2021}. For the EO investigation,  we also used a unit cell of 12 atoms (4 formula units). One should note that in the case of the r$-$phase, an instability (with negative/imaginary frequency) of TO $A_{1}$-TO$_{1}$ phonon mode was observed when the phonon was computed at the optimized unit cell volume for r-ZrO$_{2}$ (141 $\AA^{3}$) and r-HfO$_{2}$ (136 $\AA^{3}$), the optimized unit cells being given in SM \cite{Ali2022}. For the calculation of EO coefficients and other properties to be reliable, one should have only real frequencies. To achieve such a condition for ZrO$_{2}$, an experimental volume ($\sim$132 $\AA^{3}$) was imposed together with a rhombohedral angle equal to 89.56° \cite {Ali2021}, forcing the imaginary phonon mode that is calculated at 82\textit{i} $cm^{-1}$ to move to 76 $cm^{-1}$ at the experimental volume. While for HfO$_{2}$, no r$-$phase unit cell volume was experimentally reported, the volume was reduced until achieving a positive phonon frequency, and the volume used here is 128 $\AA^{3}$. 
\begin{table} [H]
    \centering
    \captionsetup{width=.6\linewidth}
     \caption{ Band gap and refractive indices of  rhombohedral R3m  ZrO$_{2}$ and HfO$_{2}$  at 633 nm. The rhombohedral polarization P$_r$ from Berry phase calculation is also shown.}
\begin{tabular}{c c c} 
 \hline
   & r-ZrO$_{2}$ & r-HfO$_{2}$\\ [1ex] 
 \hline\hline
 gap (eV) &5.79& 6.77\\  [1ex]
 n$_{xx}$ & 2.282& 2.114 \\
 n$_{zz}$ & 2.271& 2.107  \\ 
 $\delta n$ & 0.011&0.007  \\  [1ex]
 P$_{r}$ ($\mu$C/cm$^{2})$ & 6.8 & 5.1 \\ 
 \hline  \hline 
\end{tabular}
 \label{table:6}
\end{table}
The computed band gap, refractive indices, and polarization are displayed in Table \ref {table:6}. The band gaps obtained for ZrO$_2$ and HfO$_2$ in the r$-$phase are significantly higher than in the o$-$phase. The remanent polarization for the r$-$phase was also computed via Berry Phase using QE. The value obtained for the polarization is lower than in the case of the o$-$phase, as expected from previous calculations \cite {Wei2018, Silva2021}. Note that this polarization depends strongly on the compressive strain, which is directly related to the rhombohedral angle \cite{Wei2018, Ali2021}. Here, the polarization is computed at a fixed angle using the rhombohedral primitive unit cell. In contrast, in Refs.\cite {Wei2018, Silva2021} the polarization is computed with hexagonal unit cells and at different compressive strains. 
\begin{table}[H]
    \centering
    \captionsetup{width=.8\linewidth}
     \caption{The four independent elements  of $r_{ijk}(-\omega_{\sigma}; \omega, 0)$ tensor for R3m space group, together with their respective electronic, vibrational and piezoelectric contributions, computed at a wavelength of 633 nm. All values are given in pm/V.}
\begin{tabular}{c c c c c c c c c c}
\hline
\multirow{5}{1.5cm} &  & \multicolumn{5}{p{2.5cm}}{\centering r-ZrO$_2$}  &  \multicolumn{2}{p{2.5cm}}{\centering r-HfO$_2$} \\   [1ex] 
  & \multicolumn{1}{c}{r$^{ele}_{ijk}$ } & \multicolumn{1}{c}{r$^{vib}_{ijk}$ } & \multicolumn{1}{c}{r$^{p}_{ijk}$ } & \multicolumn{1}{c}{r$_{ijk}$} & & \multicolumn{1}{c}{r$^{ele}_{ijk}$ } & \multicolumn{1}{c}{r$^{vib}_{ijk}$ } & \multicolumn{1}{c}{r$^{p}_{ijk}$ } & \multicolumn{1}{c}{r$_{ijk}$}\\   [1ex] 
 r$_{33}$ &- 0.11 &  - 0.30 & -0.01 & -0.42& & -0.07 & -0.21  &-0.02&-0.30 \\  
 r$_{13}$ & 0.05 & 0.14& 0.05 & 0.24& & 0.03 &0.13&0.11 &  0.27\\   
r$_{11}$ & -0.06 &  -0.21&  -0.20& -0.28& & -0.05 &  -0.08 & -0.16& 0.29 \\ 
r$_{51}$ &0.05 &0.08&  -0.25  &-0.12 & & 0.03 & 0.21  &-0.19 & 0.05\\  
\hline  \hline 
\end{tabular}
\label{table:7}
\end{table}
As seen for o$-$phase (section \ref{ss:o-ZHO}), a unit cell with 12 atoms results in 36 vibrational modes, 3 for translation modes, and 33 for optical phonon modes. In the case of R3m space group, there are three different modes $A_{1}$, $A_{2}$, and $E$; the irreducible representation at $\Gamma$ zone center  is:
\begin{equation}\label{eq:12}
 \Gamma = 8 A_{1} + 3 A_{2} + 11  E
\end{equation}
The $A_{1}$ and (degenerate) $E$ modes are simultaneously Raman and IR active, while the A$_{2}$ modes are Raman active and IR inactive. The calculated $A_{1}$ mode frequencies are shown in Table \ref{table:4} for ZrO$_2$ and HfO$_2$, experimentally no data are available for ZrO$_2$-HfO$_2$ in r$-$phase. The $E$ and $A_{2}$ phonon frequencies are given in SM (Table S.8 in SM \cite{Ali2022})\\
Following the same method as described in sections \ref {ss:LiNbO3} and \ref {ss:o-ZHO}, the EO coefficients obtained for ZrO$_{2}$ and HfO$_{2}$ r$-$phase are given in Table \ref {table:7}. Comparing these EO coefficients to the ones of LNO and  o$-$phase, r$-$phase shows very modest EO values. This trend is also observed in terms of remanent polarization obtained for r$-$phase compared to orthorhombic Pbc2$_{1}$ (Tables \ref{table:3} and \ref{table:6}).  

\begin{figure} [H]
    \centering
     \captionsetup{width=.5\linewidth}
     \hspace*{-0.2in}
    \includegraphics [scale=0.5]{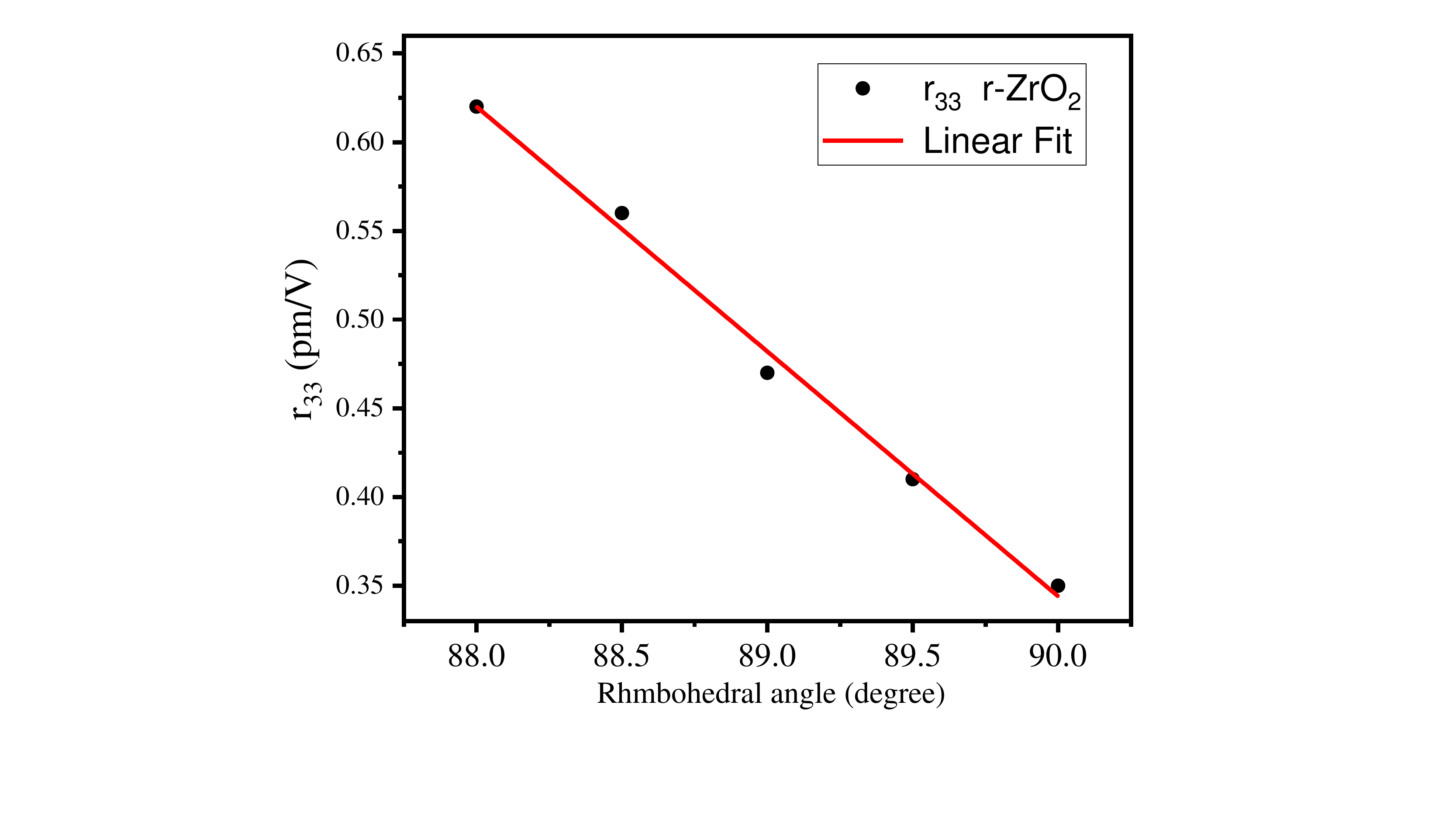}
     \vspace*{-12mm}
    \caption{r$_{33}$ EO coefficient of r- ZrO$_2$  computed at different angles.}
    \label{fig:4}
\end{figure}
\textit{Strain dependence of EO coefficients}: Because the ZrO$_2$-HfO$_2$ R3m phase has a strong dependence on the compressive strain, r$_{33}$ EO coefficient of r-ZrO$_2$ is computed at different rhombohedral angles. The results are displayed in Figure \ref {fig:4}, showing a linear dependence of r$_{33}$ with respect to the rhombohedral angle. Even at a high compressive strain (rhombohedral angle of 88°), the EO coefficient is still lower than the one calculated for o-ZrO$_2$. Two main reasons cause the low EO coefficients of r-ZrO$_2$ compared to o-ZrO$_2$. First is the relatively high band gap in the r$-$phase, which reduces the electronic contribution. The second reason is that the $A_1$ low-frequency phonon modes have negligible contributions to EO coefficients in the r$-$phase, while similar modes highly contribute to the ones of the o$-$phase (see section \ref{ss:o-ZHO}). In more details, for the r$-$phase, from 8 $A_1$ active modes (Table \ref {table:4}), only 3 modes at 220, 352, and 449 $cm^{-1}$ contribute significantly to r$_{33}$ and  r$_{13}$. The modes at 220 and 352 $cm^{-1}$ give almost the same value of the hyper-polarizabilities $\beta_{33}$ (equation \ref{eq:6}), but with opposite signs, hence they cancel each other, therefore solely the mode at 449 $cm^{-1}$ is contributing to r$_{33}$. This is the most probable reason why the r$-$phase shows a lower EO coefficient than the o$-$phase. Finally, this gives r$-$phase a very low figure of merit compared to o$-$phase and to LNO as illustrated in Figure \ref{fig:3}.

\section {Conclusion}
In conclusion, we investigated the electro-optic Pockels coefficients of the ZrO$_2$ and HfO$_2$ compounds in two noncentrosymmetric phases: rhombohedral R3m and orthorhombic Pbc2$_{1}$. We determined that the o$-$phase has considerably higher EO coefficients than the r$-$phase. This remains true for a highly strained rhombohedral phase; the optical nonlinearity is weak compared to the o$-$phase. Moreover, r$_{33}$ and r$_{13}$ in the orthorhombic phase are higher in ZrO$_2$ than in HfO$_2$, with values of about 12 and 7 pm/V  for ZrO$_2$ and HfO$_2$, respectively. Comparing the EO coefficients of o-ZrO$_2$ to rhombohedral LiNbO$_3$ reference material, the r$_{13}$ coefficient shows a comparable value to the r$_{13}$ of LiNbO$_3$. In contrast, the r$_{33}$ coefficient of o-ZrO$_2$ is three times lower than the r$_{33}$ of LiNbO$_3$. We observed that the polar and low-frequency phonon modes contribute most to the EO in the orthorhombic phase; the same trend is also observed in LiNbO$_3$. Our numerical results suggest that the ferroelectric orthorhombic phase of ZrO$_2$ and HfO$_2$-based thin films can be relevant for EO applications when considering their compatibility with silicon. Finally, considering the dominant contribution by the ionic part as identified in this work, we believe that doping ZrO$_2$/HfO$_2$ with a suitable element to stabilize the orthorhombic phase but also increase the ionicity  of these materials will lead to high EO coefficients. Thus, ZrO$_2$/HfO$_2$-based compounds may be promising candidates for EO properties and their use in photonic devices.
\\
\\
\textbf{Acknowledgements:} This work has received support from the Agence nationale de la recherche (ANR) under project FOIST (N°ANR-18-CE24-0030), and from the French national network RENATECH for nanofabrication.

\medskip
\pagebreak

\setcounter{secnumdepth}{0}
\section{SUPPLEMENTAL MATERIAL}
\setcounter{equation}{0}
\setcounter{table}{0}
\setcounter{figure}{0}
\section{EO coefficients of LiNbO$_3$ and ZrO$_2$-HfO$_2$ materials} \label{s:1}

The electro-optic (EO) coefficients for 3m  space group (SG) \cite {modernphotonics}, thus for ZrO$_{2}$-HfO$_{2}$  rhombohedral R3m (SG N° 160) and LiNbO$_{3}$  R3c (SG N° 161), have 4 independent EO coefficients with two possible representation:
\begin{align}\label{eq:13}
 m\perp x_{1}\begin{vmatrix} 
r_{11} & 0 & r_{13} \\
-r_{11} & 0 & r_{13} \\
0 & 0 & r_{33} \\
0 & r_{51} & 0 \\
r_{51} & 0 & 0 \\
0 &- r_{11} & 0 \\
\end{vmatrix} & & 
&or& m\perp x_{2}\begin{vmatrix} 
0 & -r_{22} & r_{13} \\
0 & r_{22} & r_{13} \\
0 & 0 & r_{33} \\
0 & r_{51} & 0 \\
r_{51} & 0 & 0 \\
-r_{22} & 0 & 0 \\
\end{vmatrix}
\end{align}
In the case of Pbc2$_1$ (SG N° 29 and 2mm point group), there are five independent electro-optic coefficients:
\begin{align}\label{eq:14}
\begin{vmatrix} 
0 & 0 & r_{13} \\
0 & 0 & r_{23} \\
0 & 0 & r_{33} \\
0 & r_{42} & 0 \\
r_{51} & 0 & 0 \\
0 &0 & 0 \\
\end{vmatrix}
\end{align}
\section{ Symmetry of $\beta$ (SHG) and $\beta$ (EO / dc-Pockels)  }
In the general case, the second order non-linearity $\beta_{ijk}$ is written as
\begin{equation}\label{eq:15}
\beta_{ijk}  =  \beta_{ijk} (-\omega_{\sigma}; \omega_{1}, \omega_{2})
\end{equation}
\textbf {Second harmonic generation (SHG)}\\
For SHG,  $\beta_{ijk}$  ($\omega_{\sigma}$ = 2 $\omega$ and  $\omega_{1}$ = $\omega_{2}$ = $\omega $ )     can be written as
\begin{equation}\label{eq:16}
  \beta_{ijk} = \beta_{ijk} (-2\omega; \omega, \omega)    
\end{equation}
So in the case of second harmonic generation, for example with $\beta_{yxy}$ equal to $\beta_{yyx}$, in general the symmetry given in (\ref{eq:16}) can simply be written as $\beta_{ijk} (-2\omega; \omega, \omega) =\beta_{ikj} (-2\omega; \omega, \omega) $.\\
\\
\textbf{EO / dc-Pockels effect}\\
For dc-Pockels effect, $\beta_{ijk}$  ($\omega_{\sigma}$ =  $\omega$ , $\omega_{1}$ = $\omega$, and   $\omega_{2}$ = 0) is written as
\begin{equation}\label{eq:17}
  \beta_{ijk} = \beta_{ijk} (-\omega; \omega, 0)    =   \beta_{ikj} (-\omega; 0, \omega) 
\end{equation}
The same permutation as given above for SHG then leads to $\beta_{xxz}$ (-$\omega$; $\omega$,0) =  $\beta_{xzx}$(-$\omega$; 0, $\omega$). In other words, $\beta_{xxz}$ (-$\omega$; $\omega$,0) tensor element, where the static field is along z, is equal to  $\beta_{xzx}$(-$\omega$; 0, $\omega$), so that the static electric field is always applied along z direction. But  $\beta_{xxz}$ (-$\omega$; $\omega$,0) is not equal to  $\beta_{xzx}$ (-$\omega$; $\omega$, 0) because in the  latter tensor element the static electric field is along x. Finally, this will ensure that $r_{ijk}$ coefficients respect the permutation   $r_{ijk}  = r_{jik}$.

\section{Results} \label{s:5 }
\textbf { LiNbO$_{3}$}\\
\begin{table} [H]
\centering
\parbox{0.4\textwidth}{
\begin{footnotesize}
\begin{tabular}{c c c c} 
 \hline
  LiNbO$_{3}$ & a (\AA) &b  (\AA) &c  (\AA)\\ 
       &  5.216&5.216&14.14\\
 \hline\hline
Atom & x/a& y/b& z/c \\  [1ex] 
 Li & 0& 0&0.2614 \\
Nb & 0& 0& -0.0188 \\  [1ex] 
 O& -0.0473& 0.3435&0.0441\\ [1ex] 
 \hline
 V ($\AA^{3}$) &111 \\
   \hline  \hline 
\end{tabular} 
\end{footnotesize}
\caption{The optimized geometry for LiNbO$_{3}$, on the right the LiNbO$_{3}$ R3c structure.}
\label{table:8}
}
\qquad
\begin{minipage}[c]{0.4\textwidth}%
\centering
    \includegraphics[width=1\textwidth]{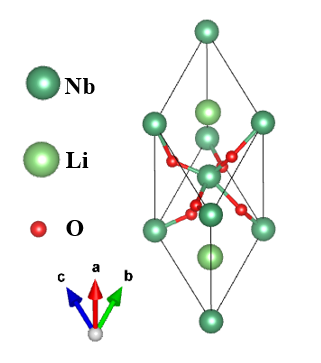}
\label{fig:figure}
\end{minipage}
\end{table}
The refractive indices of LiNbO3 calculated using B3LYP functional, together with the electronic and vibrational contributions to the polarizability $\alpha_{ij}$, are reported in Table \ref {table:2}. The ionic contribution at 633 nm is almost null.
\begin{table} [H]
    \centering
     \caption{  Refractive indices $n_{ii}$  of LiNbO$_{3}$ at 633 nm wavelength using B3LYP functional, showing the electronic and vibrational contributions to the polarizability $\alpha_{ij}$.}
\begin{tabular}{c c c | c} 
\hline
&electronic & vibrational & n$_{ii}$ \\ [1ex] 
\hline
$\alpha_{xx}$ & 244.77&  -0.60 &2.257\\[0.5ex] 
$\alpha_{yy}$ & 244.77&  -0.60 &2.257\\[0.5ex] 
$\alpha_{zz}$ & 216.36&  -0.48 &2.149\\[0.5ex] 
 \hline \hline  
\end{tabular}
 \label{table:9}
\end{table}
At the $\Gamma$ point, the optical phonons of LiNbO$_3$ can be classified according to the irreducible representation of the space group R3c into $\Gamma = 4A_{1} + 5A_{2} + 9E$. The A$_{1}$ and E modes are Raman and infrared active, while A$_{2}$ modes are inactive. 

\begin{figure} [H]
    \centering
    \includegraphics [scale=0.6] {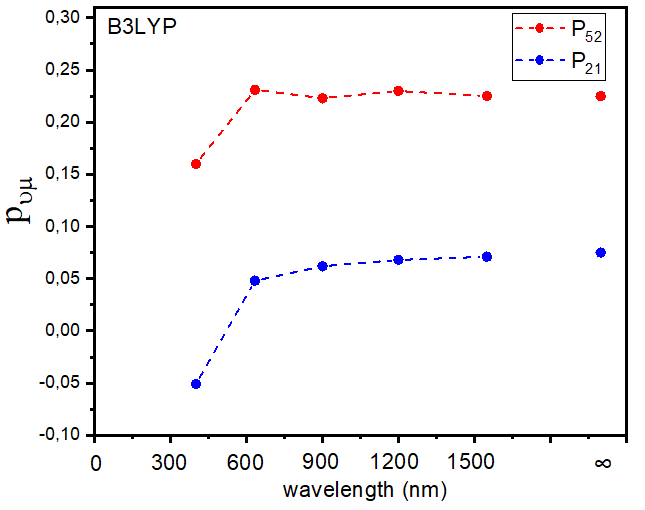}
    \caption{Independent photoelastic p$_{52}$ and p$_{21}$ coefficients of the LiNbO$_3$ crystal computed at B3LYP level as a function of the electric field wavelength (nm).}
    \label{fig:5}
\end{figure}

The calculated frequencies of the A$_{1}$ and E modes are shown in Table \ref{table:3} and compared with previous theoretical and experimental data. There is a good agreement between our calculations and available experimental values. In particular, B3LYP functional calculated frequencies are close to the experimental ones, with only a relative underestimate of the first A$_{1}$ mode. This might be due to the anharmonicity of this mode: The experimental frequency is substantially higher than the one calculated in the harmonic approximation; Considering the one-dimensional non-interacting oscillator approximation, a good approximation to the experimental value was obtained by Inbar and Cohen \cite {PRB-Inbar1996}. Increasing HF exchange (PBE0) gives relatively higher frequency values, while decreasing exchange (PBE(10$\%$) and PBE) brings lower values compared to the experimental results.\\
\begin{table} [H]
    \centering
     \caption{ Raman and IR active phonon mode frequencies (cm$^{-1}$) of ferroelectric LiNbO$_{3}$ calculated with several functionals, compared with previous theoretical and experimental results.}
\begin{tabular}{c c c c c c c c} 
 \hline
   & B3LYB &PBE0 &PBE(10$\%$)& PBE & LDA \cite {PRB2000}  & exp \cite {Kojima_1993} & exp \cite {Claus1972}\\ [1ex] 
 \hline\hline
 A1 & 232 &251&242&234&208 & 256 & /\\ 
   & 275 &291&279& 271&279 & 275&/\\

   & 340&346&332&322 &344 &332 &/\\
   
   & 636 &650&628& 611&583 & 637 &/\\
 \hline
 E & 143 &152&144& 140 &151 &  /& 155\\ 

  & 233 &241&229&221 &/&  /& 238\\

 &246 &259&250 &245&236 &   /& 265\\

 & 304 &315&302&294&307 &  / & 325\\
  & 344 &367&252&343& 334&  / & /\\
   & 361 &380&365& 355&352& / &/\\
    & 427 &434&414& 402&432 & /  &431\\
     & 576 &589&572&560 &526 &/   &582\\
     & 655 &674&654& 641&617 & / &668\\      
 \hline  \hline 
\end{tabular}
 \label{table:10}
\end{table}
In order to obtain the EO coefficients of LiNbO$_{3}$ using CRYSTAL17, we computed the electronic and vibrational contributions to the second hyperpolarizability $\beta_{ijk}(-\omega_{\sigma}; \omega, 0)$ at a wavelength of 633 nm as described in the main paper, then deduced the $\chi^{(2)}_{ijk}(-\omega_{\sigma}; \omega, 0)$ values that are given in Table  \ref{table:4} (see the main paper for the relation between the susceptibility $\chi^{(2)}_{ijk}$ and the hyperpolarizability $\beta_{ijk}$). For symmetry reasons, only 11 out of 27 elements of the rank three $\chi^{(2)}_{ijk}$ tensor are not null and only four elements are independent. These four elements are given in Table \ref{table:4}. We note that pure PBE functional gives the highest values of $\chi^{(2)}_{ijk}$, that is because pure DFT-PBE underestimates the gap, as discussed in the main paper. 
\begin{table}[H]
    \centering
     \caption{Non-zero elements of $\chi^{(2)}_{ijk}(-\omega_{\sigma}; \omega, 0)$ (pm/V) tensor in LiNbO$_{3}$ computed at a wavelength of 633 nm using different functionals, with their vibrational and electronic contributions.}
\begin{tabular}{c c c c c c c c c} 
\hline
\multirow{5}{1cm} & \multicolumn{2}{p{2.5cm}}{\centering B3LYP} &  \multicolumn{2}{p{2.5cm}}{\centering PBE0} &  \multicolumn{2}{p{2.5cm}}{\centering PBE(10$\%$)}&  \multicolumn{2}{p{2.5cm}}{\centering PBE}\\
 $\chi^{(2)}_{ijk}$ & \multicolumn{1}{c}{vib} & \multicolumn{1}{c}{ele} & \multicolumn{1}{c}{vib} & \multicolumn{1}{c}{ele} & \multicolumn{1}{c}{vib}& \multicolumn{1}{c}{ele}& \multicolumn{1}{c}{vib}& \multicolumn{1}{c}{ele}\\ [1ex] 
  
xxx  & 88.83 &13.28 & 66.43 & 8.18 &  104.8 & 19.53&136.17&50.30\\  
  
xxz & -103.00 & -35.50& -84.37 &  -28 &-129.2&-53.15&-183.4&-66.36\\ 
yzy & -228.52 & -36.56& -188.73 & -27.60 &-269.1&-54.53&-338.5&-102.4\\   
zzz & -250.08 & -76.66 &-220.04 & -63.05&-296.9&-103.1&-396.05&-117.8\\    
\hline \hline
\end{tabular}
  \label{table:11}
\end{table}
Table \ref{table:12} gives the clamped and unclamped EO coefficients of LiNbO$_3$, considering only the four independent elements of r$_{ijk}$, and compares them with previous calculations and experiments. All functionals used in the present study agree well with experimental results, notably for the r$_{33}$ and r$_{13}$ coefficients. The clamped EO coefficients, taking in consideration the piezoelectric contribution, are also given in Table \ref{table:4}.
\begin{table} [H]
    \centering
     \caption{ Total r$_{ijk}(-\omega_{\sigma}; \omega, 0)$ EO tensor (pm/V) in LiNbO$_{3}$ computed at a wavelength of 633 nm using different functionals, together with previous calculations and experiments.}
\begin{tabular}{c c c c c c c c c c} 
 \hline
  &  & B3LYB & PBE0 & PBE(10$\%$)& PBE& LDA \cite {PRBVeithen2004} &  exp \cite{Turner1966} & exp \cite{K.K.WONG2002} \\ 
\hline\hline
clamped&r$_{33}$ & 30.64 &   28.02& 30.45 &30.87& 26.9   & 30.8  &34\\  
&r$_{13}$ & 10.68 & 9.14&   11.39 & 12.84& 9.7  & 8.6 &10.9\\  
&r$_{11}$ & -7.86 &    -6.06& -7.79 &-9.59& 4.6  & 3.4 &/\\  
&r$_{51}$ & 22.51 &  19.41&  21.99 &24.51& 14.9 &  28&/\\  
\\
unclamped & & & &  & &  & &exp \cite{Abdi1998}\\
& r$_{33}$ & 31.59 &   34.32& 30.90 &31.20&  27   &32.6 & /\\  
&r$_{13}$ & 12.10 & 10.92&   12.61  &13.81&  10.5  &10& /\\  
&r$_{11}$ & -7.00 &    -5.00& -6.84 &-10.20&   7.5  &6& 9.9\\  
&r$_{51}$ & 33.28 &  27.09& 31.94 &34.04&  28.6& 32.2& /\\  
\hline  \hline
\end{tabular}
\label{table:12}
\end{table}
As the experimental values of EO coefficients are usually measured at finite electric field wavelength, in Figure \ref{fig:5} we show the dependence of photoelastic coefficients on the electric field wavelength (nm). From 633 nm up to the near-infrared wavelength range, the two coefficients remain almost constant, while below 633 nm a decrease is observed.

\textbf { ZrO$_{2}$-HfO$_{2}$}\\

\begin{table} [H]
    \centering
     \caption{ The optimized geometry for ZrO$_{2}$  orthorhombic Pbc2$_{1}$.}
\begin{tabular}{c c c c} 
 \hline
  o-ZrO$_{2}$ & a (\AA) &b  (\AA) &c  (\AA)\\ 
       &  5.34&5.14&5.16\\
 \hline\hline
Atom & x/a& y/b& z/c \\  [1ex] 
Zr& 0.0274& 0.2644&-0.2545 \\ [1ex] 
O1 & 0.3602& 0.0619& -0.1021 \\  [1ex] 
O2& 0.2262& -0.4688&-0.0064\\ [1ex] 
 \hline
 V ($\AA^{3}$) &141.62 \\
   \hline  \hline 
\end{tabular} 
\label{table:13}
\end{table}

\begin{table} [H]
    \centering
     \caption{ The optimized geometry for HfO$_{2}$  orthorhombic Pbc2$_{1}$.}
\begin{tabular}{c c c c} 
 \hline
  o-ZrO$_{2}$ & a (\AA) &b  (\AA) &c  (\AA)\\ 
       &  5.28&5.07&5.10\\
 \hline\hline
Atom & x/a& y/b& z/c \\  [1ex] 
Zr& 0.0312& 0.2651&-0.2551 \\ [1ex] 
O1 & 0.3618& 0.0638& -0.1051 \\  [1ex] 
O2& 0.2267& -0.4699&-0.00278\\ [1ex] 
 \hline
 V ($\AA^{3}$) &136.52 \\
   \hline  \hline 
\end{tabular} 
\label{table:14}
\end{table}
The orthorhombic Pbc2$_{1}$ phase with a unit of 12 atoms (4 formula units) results in 36 vibrational modes, 3 for translation modes, and 33 optical phonon modes having the following irreducible representation at the zone center $ \Gamma$: \\
$$ \Gamma = 8 A_{1} + 9 A_{2} + 8 B_{1} + 8 B_{2}$$
The computed phonon frequencies are listed in Table \ref{table:8} .\\

For the rhombohedral phase R3m, also with a unit of 12 atoms (4 formula units) resulting in 36 vibrational modes, 3 for translation modes and 33 optical phonon modes, the  irreducible representation at the zone center $ \Gamma$ is:
$$ \Gamma = 8 A_{1} + 3 A_{2} + 11  E $$
The calculated frequencies are given in Table \ref{table:8}. Note that the computed E modes here are degenerate. The lowest E-TO1 (transverse optical) mode frequency depends strongly on the unit cell volume and can also be affected by other factors, like strain, oxygen vacancies, etc.
\begin{table}[H]
    \centering
     \caption{Raman and IR phonon mode frequencies (cm$^{-1}$) of ferroelectric ZrO$_{2}$ and HfO$_{2}$ calculated using B3LYP functional. Experimental data are also reported when available.}
\begin{tabular}{c c c c c c c c c c}
\hline
\multirow{2}{1.5cm} &  \multicolumn{2}{p{2.5cm}}{\centering o-ZrO$_2$}  &  \multicolumn{1}{p{1.5cm}}{\centering o-HfO$_2$} &  \multicolumn{1}{p{1.5cm}}{\centering }&  \multicolumn{1}{p{1.5cm}}{\centering r-ZrO$_2$} &  \multicolumn{1}{p{1.5cm}}{\centering r-HfO$_2$}\\   [1ex] 
  & \multicolumn{1}{c}{B3LYP} & \multicolumn{1}{c}{exp \cite {Uwe-2022-Raman}}  & \multicolumn{1}{c}{B3LYP} &  & \multicolumn{1}{c}{B3LYP}  & \multicolumn{1}{c}{B3LYP}  \\   [1ex] 
A$_1$ & 102 &   /& 113 &A$_1$&171 &118\\  
 & 186 &200& 153& & 220&171\\  
 & 291 &  /& 253 &&267&267\\  
 & 315& 320& 295&&352&284\\  
 & 351& 340& 329& &449&423\\  
 &387&/ &369&&534&537\\
 &445&/&451&&674&692\\
 &566& 580&589&&755&770\\ [1ex] 
 A$_2$ &172 &/& 122&A$_2$&169&117\\ 
 &180&/&134&&332&312\\
 &192&200&148&&586&595\\
 &295&/&303&\\
 &400&/&414&E&76&20\\
 &482&/&496&&165&114\\
 &561&/&581&&174&123\\
 &634&/&654&&233&187\\
 &659&/&674&&325&271\\[1ex]
 B$_1$& 188& /&141&&358&304\\
 &224&/&218&&456&436\\
 &298&/&238&&536&540\\
 &329&/&311&&586&594\\
 &402&/&395&&683&702\\
 &514&/&533&&721&726\\
 &545& 550&554&\\
 &726& /&736&\\ [1ex] 
 B$_2$& 165 & /&121&\\
 & 233& /&226&\\
 & 289& /&269&\\
 &345 & /&312&\\
 & 399&/ &369&\\
 & 461& /&484&\\
 & 622 &620&637&\\
 & 683&/ &697&\\
\hline  \hline 
\end{tabular}
\label{table:15}
\end{table}

\medskip
\pagebreak
\bibliographystyle{unsrt}
\bibliography{references.bib}

\begin{thebibliography}{10}

\bibitem{Mohapatra2008}
A.~K. Mohapatra, M.~G. Bason, B.~Butscher, K.~J. Weatherill, and C.~S. Adams.
\newblock {A giant electro-optic effect using polarizable dark states}.
\newblock {\em Nature Physics}, 4(11):890--894, 2008.

\bibitem{Melikyan2014}
A.~Melikyan, L.~Alloatti, A.~Muslija, D.~Hillerkuss, P.~C. Schindler, J.~Li,
  R.~Palmer, D.~Korn, S.~Muehlbrandt, D.~{Van Thourhout}, B.~Chen, R.~Dinu,
  M.~Sommer, C.~Koos, M.~Kohl, W.~Freude, and J.~Leuthold.
\newblock {High-speed plasmonic phase modulators}.
\newblock {\em Nature Photonics}, 8(3):229--233, 2014.

\bibitem{Rueda2019}
A.~Rueda, F.~Sedlmeir, M.~Kumari, G.~Leuchs, and H.~G.~L. Schwefel.
\newblock {Resonant electro-optic frequency comb}.
\newblock {\em Nature}, 568(7752):378--381, 2019.

\bibitem{Abel2019}
S.~Abel, F.~Eltes, J.~E. Ortmann, A.~Messner, P.~Castera, T.~Wagner,
  D.~Urbonas, A.~Rosa, A.~M. Gutierrez, D.~Tulli, P.~Ma, B.~Baeuerle,
  A.~Josten, W.~Heni, D.~Caimi, L.~Czornomaz, A.~A. Demkov, J.~Leuthold,
  P.~Sanchis, and J.~Fompeyrine.
\newblock {Large Pockels effect in micro- and nanostructured barium titanate
  integrated on silicon}.
\newblock {\em Nature Materials}, 18(1):42--47, 2019.

\bibitem{OBrien2009}
J.~L O'Brien, A.~Furusawa, and J.~Vu{\v{c}}kovi{\'{c}}.
\newblock {Photonic quantum technologies}.
\newblock {\em Nature Photonics}, 3(12):687--695, 2009.

\bibitem{Wang2020}
J.~Wang, F.~Sciarrino, A.~Laing, and M.~G. Thompson.
\newblock {Integrated photonic quantum technologies}.
\newblock {\em Nature Photonics}, 14(5):273--284, 2020.

\bibitem{George2019}
J.~K. George, A.~Mehrabian, R.~Amin, J.~Meng, T.~F. de~Lima, A.~N. Tait, B.~J.
  Shastri, T.~El-Ghazawi, P.~R. Prucnal, and V.~J. J.~Sorger.
\newblock Neuromorphic photonics with electro-absorption modulators.
\newblock {\em Opt. Express}, 27(4):5181--5191, Feb 2019.

\bibitem{Shen2017}
Y.~Shen, N.~C. Harris, S.~Skirlo, M.~Prabhu, T.~Baehr-Jones, M.~Hochberg,
  X.~Sun, S.~Zhao, H.~Larochelle, D.~Englund, and M.~Solja{\v{c}}i{\'{c}}.
\newblock {Deep learning with coherent nanophotonic circuits}.
\newblock {\em Nature Photonics}, 11(7):441--446, 2017.

\bibitem{Offrein2020}
B.~J. Offrein, J.~Geler-Kremer, J.~Weiss, R.~Dangel, P.~Stark, A.~Sharma,
  S.~Abel, and F.~Horst.
\newblock Prospects for photonic implementations of neuromorphic devices and
  systems.
\newblock In {\em 2020 IEEE International Electron Devices Meeting (IEDM)},
  pages 7.4.1--7.4.4, 2020.

\bibitem{Lines2001}
M.~E. Lines and A.~M. Glass.
\newblock {\em {Principles and Applications of Ferroelectrics and Related
  Materials}}.
\newblock Oxford University Press, Oxford, 2001.

\bibitem{Turner1966}
E.~H. Turner.
\newblock High-frequency electro-optic coefficients of lithium niobate.
\newblock 303(November 2004):303--305, 1966.

\bibitem{Wooten2000-reviewonLNO}
E.~L. Wooten, K.~M. Kissa, A.~Yi-Yan, E.~J. Murphy, D.~A. Lafaw, P.~F.
  Hallemeier, D.~Maack, D.~V. Attanasio, D.~J. Fritz, G.~J. McBrien, and D.~E.
  Bossi.
\newblock A review of lithium niobate modulators for fiber-optic communications
  systems.
\newblock {\em IEEE Journal of Selected Topics in Quantum Electronics},
  6(1):69--82, 2000.

\bibitem{Wang2018}
C.~Wang, M.~Zhang, X.~Chen, M.~Bertrand, A.~Shams-Ansari, S.~Chandrasekhar,
  P.~Winzer, and M.~Loncar.
\newblock {Integrated lithium niobate electro-optic modulators operating at
  CMOS-compatible voltages}.
\newblock {\em Nature}, 562(7725):101--104, 2018.

\bibitem{Rabiei2013HeterogeneousLN}
Payam R., Jichi M., Saeed K., Jeff C., and Sasan F.
\newblock Heterogeneous lithium niobate photonics on silicon substrates.
\newblock {\em Optics express}, 21 21:25573--81, 2013.

\bibitem{PRB-Zgonik1994}
M.~Zgonik, P.~Bernasconi, M.~Duelli, R.~Schlesser, P.~Gunter, M.~H. Garrett,
  D.~Rytz, Y.~Zhu, and X.~Wu.
\newblock Dielectric, elastic, piezoelectric, electro-optic, and elasto-optic
  tensors of {BaTiO$_3$} crystals.
\newblock {\em Phys. Rev. B}, 50:5941--5949, Sep 1994.

\bibitem{Messner2019}
A.~Messner, F.~Eltes, P.~Ma, S.~Abel, B.~Baeuerle, A.~Josten, W.~Heni,
  D.~Caimi, J.~Fompeyrine, and J.~Leuthold.
\newblock {Plasmonic Ferroelectric Modulators}.
\newblock {\em Journal of Lightwave Technology}, 37(2):281--290, 2019.

\bibitem{Xiong2014}
C.~Xiong, W.~H.~P. Pernice, J.~H. Ngai, J.~W. Reiner, D.~Kumah, F.~J. Walker,
  C.~H. Ahn, and H.~X. Tang.
\newblock {Active Silicon Integrated Nanophotonics: Ferroelectric BaTiO$_3$
  Devices}.
\newblock {\em Nano Letters}, 14(3):1419--1425, mar 2014.

\bibitem{Paoletta-PRB-EO}
T.~Paoletta and A.~A. Demkov.
\newblock {Pockels effect in low-temperature rhombohedral BaTiO$_{3}$}.
\newblock {\em Phys. Rev. B}, 103:014303, Jan 2021.

\bibitem{Eltes2020}
F.~Eltes, G.~E. Villarreal-Garcia, D.~Caimi, H.~Siegwart, A.~A. Gentile,
  A.~Hart, P.~Stark, G.~D. Marshall, M.~G. Thompson, J.~Barreto, J.~Fompeyrine,
  and S.~Abel.
\newblock {An integrated optical modulator operating at cryogenic
  temperatures}.
\newblock {\em Nature Materials}, 19(11):1164--1168, 2020.

\bibitem{Boscke2011}
T.~S. Böscke, J.~Müller, D.~Bräuhaus, U.~Schröder, and U.~Böttger.
\newblock Ferroelectricity in hafnium oxide thin films.
\newblock {\em Applied Physics Letters}, 99(10):102903, 2011.

\bibitem{Wei2018}
Y.~Wei, P.~Nukala, M.~Salverda, S.~Matzen, H.~J. Zhao, J.~Momand, A.~S.
  Everhardt, G.~Agnus, G.~R. Blake, P.~Lecoeur, B.~J. Kooi,
  J.~{\'{I}}{\~{n}}iguez, B.~Dkhil, and B~Noheda.
\newblock {A rhombohedral ferroelectric phase in epitaxially strained
  Hf$_{0.5}$Zr$_{0.5}$O$_2$ thin films}.
\newblock {\em Nature Materials}, 17(12):1095--1100, 2018.

\bibitem{8423435-Ali}
T.~Ali, P.~Polakowski, S.~Riedel, T.~Büttner, T.~Kämpfe, M.~Rudolph,
  B.~Pätzold, K.~Seidel, D.~Löhr, R.~Hoffmann, M.~Czernohorsky, K.~Kühnel,
  P.~Steinke, J.~Calvo, K.~Zimmermann, and J.~Müller.
\newblock {High Endurance Ferroelectric Hafnium Oxide-Based FeFET Memory
  Without Retention Penalty}.
\newblock {\em IEEE Transactions on Electron Devices}, 65(9):3769--3774, 2018.

\bibitem{doi:10.1063/1.3636417-Muller}
J.~Müller, T.~S. Böscke, D.~Bräuhaus, U.~Schröder, U.~Böttger,
  J.~Sundqvist, P.~Kücher, T.~Mikolajick, and L.~Frey.
\newblock {Ferroelectric Zr$_{0.5}$Hf$_{0.5}$O$_{2}$ thin films for nonvolatile
  memory applications}.
\newblock {\em Applied Physics Letters}, 99(11):112901, 2011.

\bibitem{Park-energystorage2014}
M.~H. Park, H.~J. Kim, Y.~J. Kim, T.~Moon, K.~Do. Kim, and C.~S. Hwang.
\newblock {Thin Hf$_x$Zr$_{1-x}$O$_2$ Films: A New Lead-Free System for
  Electrostatic Supercapacitors with Large Energy Storage Density and Robust
  Thermal Stability}.
\newblock {\em Advanced Energy Materials}, 4(16):1400610, 2014.

\bibitem{Silva-energystorage2021}
J.~P.~B. Silva, K.~C. Sekhar, H.~Pan, J.~L. MacManus-Driscoll, and M.~Pereira.
\newblock Advances in dielectric thin films for energy storage applications,
  revealing the promise of group iv binary oxides.
\newblock {\em ACS Energy Letters}, 6(6):2208--2217, 2021.

\bibitem{fork_armani-leplingard_kingston_1994}
D.~K. Fork, F.~Armani-Leplingard, and J.~J. Kingston.
\newblock Optical losses in ferroelectric oxide thin films: Is there light at
  the end of the tunnel?
\newblock {\em MRS Proceedings}, 361:155, 1994.

\bibitem{Epitaxialferroelectricoxide}
D.~Sando, Y.~Yang, C.~Paillard, B.~Dkhil, L.~Bellaiche, and V.~Nagarajan.
\newblock Epitaxial ferroelectric oxide thin films for optical applications.
\newblock {\em Applied Physics Reviews}, 5(4):041108, 2018.

\bibitem{Cheema2020}
Suraj~S Cheema, Daewoong Kwon, Nirmaan Shanker, Roberto Dos~Reis, Shang-Lin
  Hsu, Jun Xiao, Haigang Zhang, Ryan Wagner, Adhiraj Datar, Margaret~R
  McCarter, Claudy~R Serrao, Ajay~K Yadav, Golnaz Karbasian, Cheng-Hsiang Hsu,
  Ava~J Tan, Li-Chen Wang, Vishal Thakare, Xiang Zhang, Apurva Mehta, Evguenia
  Karapetrova, Rajesh~V Chopdekar, Padraic Shafer, Elke Arenholz, Chenming Hu,
  Roger Proksch, Ramamoorthy Ramesh, Jim Ciston, and Sayeef Salahuddin.
\newblock {Enhanced ferroelectricity in ultrathin films grown directly on
  silicon}.
\newblock {\em Nature}, 580(7804):478--482, 2020.

\bibitem{undopedHfO2}
P.~Polakowski and J.~Müller.
\newblock Ferroelectricity in undoped hafnium oxide.
\newblock {\em Applied Physics Letters}, 106(23):232905, 2015.

\bibitem{Starschich2017}
S.~Starschich, T.~Schenk, U.~Schroeder, and U.~Boettger.
\newblock {Ferroelectric and piezoelectric properties of
  Hf$_{1-x}$Zr$_{x}$O$_{2}$ and pure ZrO$_2$ films}.
\newblock {\em Applied Physics Letters}, 110(18):182905, 2017.

\bibitem{Shimura-2021}
R.~Shimura, T.~Mimura, A.~Tateyama, T.~Shimizu, T.~Yamada, Y.~Tanaka, Y.~Inoue,
  and H.~Funakubo.
\newblock {Preparation of 1~$\mu$m thick Y$-$doped HfO$_2$ ferroelectric films
  on (111) Pt/TiO$_x$/SiO$_2$/(001){Si} substrates by a sputtering method and
  their ferroelectric and piezoelectric properties}.
\newblock {\em Japanese Journal of Applied Physics}, 60(3):031009, mar 2021.

\bibitem{Lyu2018}
J.~Lyu, I.~Fina, R.~Solanas, J.~Fontcuberta, and F.~Sánchez.
\newblock {Robust ferroelectricity in epitaxial Hf$_{1/2}$Zr$_{1/2}$O$_2$ thin
  films}.
\newblock {\em Applied Physics Letters}.

\bibitem{o-HfO2-2021}
Peijie Jiao, Jiayi Li, Zhongnan Xi, Xiaoyu Zhang, Jian Wang, Yurong Yang,
  Yu~Deng, and Di~Wu.
\newblock {Ferroelectric Hf$_{0.5}$Zr$_{0.5}$O$_2$ thin films deposited
  epitaxially on (110)$-$oriented SrTiO$_3$}.
\newblock {\em Applied Physics Letters}, 119(25):252901, 2021.

\bibitem{Yun2022}
Y.~Yun, P.~Buragohain, M.~Li, Z.~Ahmadi, Y.~Zhang, X.~Li, H.~Wang, J.~Li,
  P.~Lu, L.~Tao, H.~Wang, J.~E. Shield, E.~Y. Tsymbal, A.~Gruverman, and X.~Xu.
\newblock {Intrinsic ferroelectricity in Y-doped HfO$_2$ thin films}.
\newblock {\em Nature Materials}, 21(8):903--909, 2022.

\bibitem{Silva2021}
J.~P.~B. Silva, R.~F. Negrea, M.~C. Istrate, S.~Dutta, H.~Aramberri,
  J.~Iniguez, F.~G. Figueiras, C.~Ghica, K.~C. Sekhar, and A.~L Kholkin.
\newblock {Wake-up Free Ferroelectric Rhombohedral Phase in Epitaxially
  Strained ZrO$_2$ Thin Films}.
\newblock {\em ACS Applied Materials and Interfaces}, 13(43):51383--51392, nov
  2021.

\bibitem{Ali2021}
A.~El~Boutaybi, T.~Maroutian, L.~Largeau, S.~Matzen, and P.~Lecoeur.
\newblock {Stabilization of the epitaxial rhombohedral ferroelectric phase in
  ZrO$_{2}$ by surface energy}.
\newblock {\em Phys. Rev. Materials}, 6:074406, Jul 2022.

\bibitem{Hoffman-negative-c}
M.~Hoffmann, M.~Pešić, K.~Chatterjee, Asif~I. Khan, S.~Salahuddin,
  S.~Slesazeck, U.~Schroeder, and T.~Mikolajick.
\newblock {Direct Observation of Negative Capacitance in Polycrystalline
  Ferroelectric HfO$_2$}.
\newblock {\em Advanced Functional Materials}, 26(47):8643--8649, 2016.

\bibitem{Hoffman2019}
M.~Hoffmann, S.~Slesazeck, T.~Mikolajick, and C.~S. Hwang.
\newblock Negative capacitance in hfo2- and zro2-based ferroelectrics.
\newblock In Uwe Schroeder, Cheol~Seong Hwang, and Hiroshi Funakubo, editors,
  {\em Ferroelectricity in Doped Hafnium Oxide: Materials, Properties and
  Devices}, Woodhead Publishing Series in Electronic and Optical Materials,
  pages 473--493. Woodhead Publishing, 2019.

\bibitem{Wei2019}
Y.~Wei, S.~Matzen, T.~Maroutian, G.~Agnus, M.~Salverda, P.~Nukala, Q.~Chen,
  J.~Ye, P.~Lecoeur, and B.~Noheda.
\newblock {Magnetic Tunnel Junctions Based on Ferroelectric
  Hf$_{0.5}$Zr$_{0.5}$O$_{2}$ Tunnel Barriers}.
\newblock {\em Phys. Rev. Applied}, 12:031001, Sep 2019.

\bibitem{Goh-2018-tj}
Y.~Goh and S.~Jeon.
\newblock {The effect of the bottom electrode on ferroelectric tunnel junctions
  based on CMOS$-$compatible HfO$_2$ }.
\newblock {\em Nanotechnology}, 29(33):335201, jun 2018.

\bibitem{Kondo_2021_1}
S.~Kondo, R.~Shimura, T.~Teranishi, A.~Kishimoto, T.~Nagasaki, H.~Funakubo, and
  T.~Yamada.
\newblock {Linear electro-optic effect in ferroelectric HfO$_2$ $-$based
  epitaxial thin films}.
\newblock {\em Japanese Journal of Applied Physics}, 60(7):070905, jun 2021.

\bibitem{Kondo_2021_2}
S.~Kondo, R.~Shimura, T.~Teranishi, A.~Kishimoto, T.~Nagasaki, H.~Funakubo, and
  T.~Yamada.
\newblock {Influence of orientation on the electro-optic effect in epitaxial
  Y-doped HfO$_2$ ferroelectric thin films}.
\newblock {\em Japanese Journal of Applied Physics}, 60({SF}):SFFB13, aug 2021.

\bibitem{crystal14-2014}
R.~Dovesi, R.~Orlando, A.~Erba, C.~M. Zicovich-Wilson, B.~Civalleri,
  S.~Casassa, L.~Maschio, M.~Ferrabone, M.~De~La~Pierre, P.~D'Arco, Y.~Noël,
  M.~Causà, M.~R{\'{e}}rat, and B.~Kirtman.
\newblock {CRYSTAL14: A program for the ab initio investigation of crystalline
  solids}.
\newblock {\em International Journal of Quantum Chemistry}, 114(19):1287--1317,
  2014.

\bibitem{CRYSTAL17}
R.~Dovesi, A.~Erba, R.~Orlando, C.~M. Zicovich-Wilson, B.~Civalleri,
  L.~Maschio, M.~R{\'{e}}rat, S.~Casassa, J.~Baima, S.~Salustro, and
  B.~Kirtman.
\newblock {Quantum-mechanical condensed matter simulations with CRYSTAL}.
\newblock {\em WIREs Computational Molecular Science}, 8(4):e1360, 2018.

\bibitem{Maschio2012}
L.~Maschio, B.~Kirtman, M.~R{\'{e}}rat, R.~Orlando, and R.~Dovesi.
\newblock {Ab initio analytical Raman intensities for periodic systems through
  a coupled perturbed Hartree-Fock/Kohn-Sham method in an atomic orbital basis.
  I. Theory}.
\newblock {\em The Journal of Chemical Physics}, 139(16):164101, 2013.

\bibitem{Maschio2012-IR}
L.~Maschio, B.~Kirtman, R.~Orlando, and M.~R{\'{e}}rat.
\newblock {Ab initio analytical infrared intensities for periodic systems
  through a coupled perturbed Hartree-Fock/Kohn-Sham method}.
\newblock {\em The Journal of Chemical Physics}, 137(20):204113, 2012.

\bibitem{Maschio2013}
L.~Maschio, B.~Kirtman, M.~R{\'{e}}rat, R.~Orlando, and R.~Dovesi.
\newblock {Comment on “Ab initio analytical infrared intensities for periodic
  systems through a coupled perturbed Hartree-Fock/Kohn-Sham method” [J.
  Chem. Phys. 137, 204113 (2012)]}.
\newblock {\em The Journal of Chemical Physics}, 139(16):167101, 2013.

\bibitem{Maschio2015}
L.~Maschio, M.~R{\'{e}}rat, B.~Kirtman, and R.~Dovesi.
\newblock {Calculation of the dynamic first electronic hyperpolarizability
  $\beta$($- \omega_{\sigma}$; $\omega_1$, $\omega_2$) of periodic systems.
  Theory, validation, and application to multi-layer MoS$_2$}.
\newblock {\em The Journal of Chemical Physics}, 143(24):244102, 2015.

\bibitem{Rerat-2016}
M.~R{\'{e}}rat, L.~Maschio, B.~Kirtman, B.~Civalleri, and R.~Dovesi.
\newblock {Computation of Second Harmonic Generation for Crystalline Urea and
  KDP. An ab Initio Approach through the Coupled Perturbed
  Hartree–Fock/Kohn–Sham Scheme}.
\newblock {\em Journal of Chemical Theory and Computation}, 12(1):107--113,
  2016.
\newblock PMID: 26636615.

\bibitem{Bishop1990}
David~M. Bishop.
\newblock Molecular vibrational and rotational motion in static and dynamic
  electric fields.
\newblock {\em Rev. Mod. Phys.}, 62:343--374, Apr 1990.

\bibitem{Ebra2015}
A.~Erba, M.~T. Ruggiero, T.~M. Korter, and R.~Dovesi.
\newblock {Piezo-optic tensor of crystals from quantum-mechanical
  calculations}.
\newblock {\em The Journal of Chemical Physics}, 143(14):144504, 2015.

\bibitem{Zgonik2002}
M~Jazbin{\v{s}}ek and M.~Zgonik.
\newblock {Material tensor parameters of LiNbO$_3$ relevant for electro- and
  elasto-optics}.
\newblock {\em Applied Physics B}, 74(4):407--414, 2002.

\bibitem{PRB-Ebra2013}
A.~Erba and R.~Dovesi.
\newblock {Photoelasticity of crystals from theoretical simulations}.
\newblock {\em Phys. Rev. B}, 88:045121, Jul 2013.

\bibitem{Pascale2004}
F.~Pascale, C.~M. Zicovich-Wilson, F.~López~Gejo, B.~Civalleri, R.~Orlando,
  and R.~Dovesi.
\newblock The calculation of the vibrational frequencies of crystalline
  compounds and its implementation in the crystal code.
\newblock {\em Journal of Computational Chemistry}, 25(6):888--897, 2004.

\bibitem{Zicovich2004}
C.~M. Zicovich-Wilson, F.~Pascale, C.~Roetti, V.~R. Saunders, R.~Orlando, and
  R.~Dovesi.
\newblock {Calculation of the vibration frequencies of $\alpha$-quartz: The
  effect of Hamiltonian and basis set}.
\newblock {\em Journal of Computational Chemistry}, 25(15):1873--1881, 2004.

\bibitem{Becke1993}
A.~D. Becke.
\newblock {Density‐functional thermochemistry. III. The role of exact
  exchange}.
\newblock {\em The Journal of Chemical Physics}, 98(7):5648--5652, 1993.

\bibitem{Lee1988}
C.~Lee, W.~Yang, and R.~G. Parr.
\newblock Development of the colle-salvetti correlation-energy formula into a
  functional of the electron density.
\newblock {\em Phys. Rev. B}, 37:785--789, Jan 1988.

\bibitem{PBE1996}
J.~P. Perdew, K.~Burke, and M.~Ernzerhof.
\newblock Generalized gradient approximation made simple.
\newblock {\em Phys. Rev. Lett.}, 77:3865--3868, Oct 1996.

\bibitem{Adamo1999}
C.~Adamo and V.~Barone.
\newblock {Toward reliable density functional methods without adjustable
  parameters: The PBE0 model}.
\newblock {\em The Journal of Chemical Physics}, 110(13):6158--6170, 1999.

\bibitem{PRB-basisset-Nb}
S.~Dall'Olio, R.~Dovesi, and R.~Resta.
\newblock Spontaneous polarization as a berry phase of the hartree-fock wave
  function: The case of {KNbO}$_{3}$.
\newblock {\em Phys. Rev. B}, 56:10105--10114, Oct 1997.

\bibitem{Oliveira-basisset-Li}
D.~Vilela~Oliveira, J.~Laun, M.~F. Peintinger, and T.~Bredow.
\newblock {BSSE-correction scheme for consistent gaussian basis sets of double-
  and triple-zeta valence with polarization quality for solid-state
  calculations}.
\newblock {\em Journal of Computational Chemistry}, 40(27):2364--2376, 2019.

\bibitem{PRB-basisset-O}
R.~P. McCall, E.~M. Scherr, A.~G. MacDiarmid, and A.~J. Epstein.
\newblock {Anisotropic optical properties of an oriented-emeraldine-base
  polymer and an emeraldine-hydrochloride-salt polymer}.
\newblock {\em Phys. Rev. B}, 50:5094--5100, Aug 1994.

\bibitem{Valenzano2011}
L.~Valenzano, B.~Civalleri, S.~Chavan, S.~Bordiga, M.~H. Nilsen, S.~Jakobsen,
  K.~P. Lillerud, and C.~Lamberti.
\newblock {Disclosing the Complex Structure of UiO-66 Metal-Organic Framework:
  A Synergic Combination of Experiment and Theory}.
\newblock {\em Chemistry of Materials}, 23(7):1700--1718, apr 2011.

\bibitem{Basisset-O-HfO2}
Thomas Bredow, Karl Jug, and Robert~A. Evarestov.
\newblock {Electronic and magnetic structure of ScMnO$_3$}.
\newblock {\em physica status solidi (b)}, 243(2):R10--R12, 2006.

\bibitem{PRB-basisset-Hf}
D.~Munoz~Ramo, J.~L. Gavartin, A.~L. Shluger, and G.~Bersuker.
\newblock {Spectroscopic properties of oxygen vacancies in monoclinic HfO$_{2}$
  calculated with periodic and embedded cluster density functional theory}.
\newblock {\em Phys. Rev. B}, 75:205336, May 2007.

\bibitem{Ali2022}
{\em See Supplemental Material at [] for more details on EO and second harmonic
  generation symmetry and complementary results on phonon modes of LNO and
  ZrO$_2$-HFO$_2$, crystal structure, and atomic positions.}

\bibitem{GapLNO-1990}
A.~Dhar and A.~Mansingh.
\newblock Optical properties of reduced lithium niobate single crystals.
\newblock {\em Journal of Applied Physics}, 68(11):5804--5809, 1990.

\bibitem{Thierfelde-gap2009}
C.~Thierfelder, S.~Sanna, Arno Schindlmayr, and W.~G. Schmidt.
\newblock {Do we know the band gap of lithium niobate?}
\newblock {\em Physica Status Solidi c}, 7(2):362--365, 2010.

\bibitem{PRB-Nahm2008}
H.~H. Nahm and C.~H. Park.
\newblock {First-principles study of microscopic properties of the Nb antisite
  in LiNbO$_{3}$: Comparison to phenomenological polaron theory}.
\newblock {\em Phys. Rev. B}, 78:184108, Nov 2008.

\bibitem{Schlarb1993}
U.~Schlarb and K.~Betzler.
\newblock {Refractive indices of lithium niobate as a function of wavelength
  and composition}.
\newblock {\em Journal of Applied Physics}, 73(7):3472--3476, 1993.

\bibitem{Bergman1968}
J.~G. Bergman, A.~Ashkin, A.~A. Ballman, J.~M. Dziedzic, H.~J. Levinstein, and
  R.~G. Smith.
\newblock {Curie temperature, birefringence, and phase-matching temperature
  variations in LiNbO$_3$ as function of melt stoichiometry.}
\newblock {\em Applied Physics Letters}, 12(3):92--94, 1968.

\bibitem{PRBVeithen2004}
M.~Veithen, X.~Gonze, and P.~Ghosez.
\newblock First-principles study of the electro-optic effect in ferroelectric
  oxides.
\newblock {\em Phys. Rev. Lett.}, 93:187401, Oct 2004.

\bibitem{Nelson1974}
D.~F. Nelson and R.~M. Mikulyak.
\newblock {Refractive indices of congruently melting lithium niobate}.
\newblock {\em Journal of Applied Physics}, 45(8):3688--3689, 1974.

\bibitem{K.K.WONG2002}
{K. K. WONG}.
\newblock {Properties of Lithium Niobate}.
\newblock (ISBN 0 85296 799 3):136, 2002.

\bibitem{Abdi1998}
F.~Abdi, M.~Aillerie, P.~Bourson, M.~D. Fontana, and K.~Polgar.
\newblock {Electro-optic properties in pure LiNbO$_3$ crystals from the
  congruent to the stoichiometric composition}.
\newblock {\em Journal of Applied Physics}, 84(4):2251--2254, 1998.

\bibitem{Giannozzi2009}
P.~Giannozzi, S.~Baroni, N.~Bonini, M.~Calandra, R.~Car, C.~Cavazzoni,
  D.~Ceresoli, G.~L. Chiarotti, M.~Cococcioni, and I.~Dabo.
\newblock \textit {et al}., {QUANTUM ESPRESSO: A modular and open-source
  software project for quantum simulations of materials}.
\newblock {\em Journal of Physics Condensed Matter}, 21(39), 2009.

\bibitem{Materlik2015}
R.~Materlik, C.~Kunneth, and A.~Kersch.
\newblock {The origin of ferroelectricity in Hf$_{1-x}$Zr$_x$O$_2$: A
  computational investigation and a surface energy model}.
\newblock {\em Journal of Applied Physics}, 117(13), 2015.

\bibitem{Uwe-2022-Raman}
M.~Materano, P.~Reinig, A.~Kersch, M.~Popov, M.~Deluca, T.~Mikolajick,
  U.~Boettger, and U.~Schroeder.
\newblock {Raman Spectroscopy as a Key Method to Distinguish the Ferroelectric
  Orthorhombic Phase in Thin ZrO$_{2}$ Based Films}.
\newblock {\em Physica Status Solidi (RRL) – Rapid Research Letters},
  16(4):2100589, 2022.

\bibitem{modernphotonics}
A.~Yariv and P.~Yeh.
\newblock {\em {Photonics: Optical electronics in modern communications}}.
\newblock Oxford University Press, 6 edition, 2007.

\bibitem{Kisi1998}
E.~H. Kisi and C.~J. Howard.
\newblock {Crystal structures of zirconia phases and their inter-relation}.
\newblock {\em Key Engineering Materials}, 154(153-154):1--36, 1998.

\bibitem{Kisi}
E.~H. Kisi.
\newblock Influence of hydrostatic pressure on the t→o transformation in
  mg-psz studied by in situ neutron diffraction.
\newblock {\em Journal of the American Ceramic Society}, 81(3):741--745, 1998.

\bibitem{PRB-Inbar1996}
Iris Inbar and R.~E. Cohen.
\newblock {Comparison of the electronic structures and energetics of
  ferroelectric LiNbO$_{3}$ and LiTaO$_{3}$}.
\newblock {\em Phys. Rev. B}, 53:1193--1204, Jan 1996.

\bibitem{PRB2000}
V.~Caciuc, A.~V. Postnikov, and G.~Borstel.
\newblock {Ab initio structure and zone-center phonons in LiNbO$_{3}$}.
\newblock {\em Phys. Rev. B}, 61:8806--8813, Apr 2000.

\bibitem{Kojima_1993}
Seiji Kojima.
\newblock Composition variation of optical phonon damping in lithium niobate
  crystals.
\newblock {\em Japanese Journal of Applied Physics}, 32(Part 1, No.
  9B):4373--4376, sep 1993.

\bibitem{Claus1972}
R.~Claus, J.~Brandmüller, G.~Borstel, E.~Wiesendanger, and L.~Steffan.
\newblock {Directional Dispersion and Assignment of Optical Phonons in
  LiNbO$_3$}.
\newblock {\em Zeitschrift für Naturforschung A}, 27(8-9):1187--1192, 1972.

\end{thebibliography}
\end{document}